\documentclass[intlimits,sumlimits,12pt]{iopart}

\newcommand{\diag}{\textrm{diag}}
\newcommand{\dd}{\text{d}}
\newcommand{\trans}{\text{T}}
\newcommand{\ytens}{\left(Y\otimes \mathds{1}_N\right)}
\newcommand{\ytenstrans}{\left(Y^\trans \otimes \mathds{1}_N\right)}
\newcommand{\bralign}[1]{\phantom{#1} \quad}
\newcommand{\EQref}[1]{(\ref{#1})}

\usepackage{iopams}
\expandafter\let\csname equation*\endcsname\relax

\expandafter\let\csname endequation*\endcsname\relax

\usepackage{amsmath}
\usepackage{hyperref}
\usepackage{dsfont}
\usepackage{cite}
\usepackage{graphicx}
\usepackage{amssymb}
\usepackage{mathtools}
\usepackage[capitalise]{cleveref}
\crefname{section}{Sec.}{Secs.}
\usepackage{braket}
\usepackage[disable]{todonotes}
\usepackage{float}
\usepackage{subfig}
\usepackage{nicematrix}
\usepackage{multirow}
\usepackage{bm}

\NiceMatrixOptions{cell-space-limits = 2pt}

\DeclareMathOperator{\str}{str}
\DeclareMathOperator{\sdet}{sdet}
\DeclareMathOperator{\ord}{ord}

\newcommand*{\myprime}{^{\prime}\mkern-1.2mu}

\begin{document}

\title[]{Distribution of Scattering Matrix Elements in the case of Symplectic Symmetry}

\author{Nils Gluth and Thomas Guhr}

\address{Fakult\"at f\"ur Physik, Universit\"at Duisburg--Essen, Duisburg, Germany}
\ead{nils.gluth@uni-due.de, thomas.guhr@uni-due.de}
\vspace{10pt}

\begin{abstract}
  Scattering theory is at the core of gaining information about most quantum systems. In general, analytic progress is difficult and the investigation of universal features using statistical methods is of interest. Commonly this is carried out within the Heidelberg approach which models the Hamiltonian of the target by a random matrix. Hitherto, such investigations were performed for systems that are effectively spinless or have total angular momentum conservation. In light of recent experimental progress it has become relevant to also investigate systems with half-integer spin.
  Using supersymmetry, we derive the distribution for the real and imaginary parts of the scattering matrix elements and the corresponding cross sections in the case of symplectic symmetry. These results complete the derivations for Dysons threefold way.
\end{abstract}

\vspace{2pc}

\submitto{}

\section{Introduction}
\label{sec0}
Scattering experiments are indispensable to investigate quantum systems. Originally founded in nuclear physics \cite{Ericson1963,Satchler1963,Brink1963,Porter1965,MW1969,Feshbach1992,Zelevinsky1996,GMGW1998,PS2002,Weidenmueller2009} scattering theory has become increasingly relevant also in many other areas of research. Among these are for example condensed matter \cite{Lee1987,Jalabert1990,Beenakker1997,Alhassid2000}, classical wave systems \cite{Weaver1989,Ellegaard1995,Gros2014}, quantum graphs \cite{KS2000,GA2004,PW2013}, wireless communication \cite{CD2011} and many more \cite{FKS2005,YOAA2011,MRW2010}.

The information on the scattering process is encoded in the scattering matrix which maps incoming and outgoing asymptotic waves onto each other. For asymptotic states to exist it is vital that interaction is spatially localized. The interaction in this target region is described by a Hamiltonian $H$. Typically, the exact nature of the asymptotic states is not relevant and rather they are grouped into channels depending on, for example, wave type, amplitude or phase in quantum systems but in general are arbitrary. For a finite amount of channels the scattering matrix, whose entries are the transition amplitudes from one channel to another, can be represented as a finite dimensional unitary matrix. The scattering matrix has to be unitary to guarantee flux conservation. In such a setting the scattering matrix can be, within reasonable assumptions, expressed in terms of a resolvent as a function of the Hamiltonian $H$ \cite{MW1969}. 

To facilitate a statistical description the Hamiltonian is modelled by random matrix. This is the Heidelberg approach \cite{AWM1975} which incorporates the scattering process on a microscopic level. In prior works the statistics of the scattering matrix have been investigated for systems whose spectral statistics follow the Gaussian Unitary Ensemble (GUE), \textit{i.e.} systems without time reversal invariance, and those who follow the Gaussian Orthogonal Ensemble (GOE), \textit{i.e.} systems with integer/no spin or spin rotation symmetry. In the present work we focus on systems with half-integer spin and no additional symmetries. Such systems have recently gained increasing interested due to their possible realization via quantum graphs \cite{JMS2014} or gold nanonparticles \cite{KBSR2008}. The corresponding random matrix ensemble is the well known Gaussian Symplectic Ensemble (GSE). In two recent publications \cite{CGKGD2025,GDG2025} we have shed light on possible experimental realizations for measuring the off-diagonal scattering elements and also presented the exact distribution for real and imaginary part of the scattering matrix elements and of their corresponding cross sections. Here, we present a comprehensive derivation of these results involving many technical aspects that deserve a detached discussion. We use the novel work for the other two ensembles and derive a nonlinear sigma model for the characteristic function which we obtain by using a new variant of the supersymmetry method \cite{NKSG2014,KNSGDMRS2013} which differs from the one for correlation functions.

The exact calculation of the characteristic function unveils parameter dependencies and not only allows for comparison with and prediction of experiments but also for asymptotic treatment. In a recent work \cite{KDG2025} it was possible to solve the long outstanding problem of analytic derivation of Gaussian behaviour in the Ericson regime \cite{Ericson1960} for unitary symmetry. Our results make it possible to determine the Ericson limit in symplectic systems as well. These calculations are part of ongoing research and will be part of a forthcoming publication.

The paper is structured as follows. In \Cref{sec:scattermat} we briefly sketch salient features of scattering theory and our random matrix model. Furthermore, we list some properties that are specific to the symplectic nature of the GSE. Then we derive the exact results for the distribution of scattering matrix elements by means of supersymmetry in \cref{sec:calcdistr}. In \cref{sec:scattercross} we obtain the distribution of scattering cross sections. In \cref{sec:numerics} we numerically evaluate our results and compare to Monte-Carlo simulations. Furthermore, we shed some light onto the similarities of the systems with symplectic and orthogonal symmetry in \cref{sec:connectiongoegse}. Finally, we present a brief discussion in \cref{sec:conclusion}

\section{Scattering Matrix and Heidelberg approach}
\label{sec:scattermat}
The scattering matrix $S$ of a system with $M$ open channels and $N\times N$ target Hamiltonian $H$ under reasonable assumptions is given by \cite{MW1969,fyodorov_statistics_1997}
\begin{equation}
	S_{ab}(E) = \delta_{ab} - i 2\pi W_a^\dagger G(E) W_b
\end{equation}
with the matrix resolvent 
\begin{equation}
	G^{-1}(E) = \mathds{1}_{\alpha N} - H + i \pi \sum_{c=1}^{M} W_c W_c^\dagger.
\end{equation}
Where we define $\alpha=1$ for $\beta=1,2$ and $\alpha=2$ for $\beta=4$. The entries of $W_c$ are defined by the ensemble of choice. Here, we want to focus on the case $\beta=4$. Therefore, the $W_c$ are vectors whose entries are real quaternions, and similarly $H$ is a real quaternion self-dual matrix \cite{beenakker_random-matrix_1997} doubling its dimension. Furthermore, we choose the average scattering matrix $\langle S\rangle$ to be diagonal which corresponds to the orthogonality relation \cite{VWZ1985}
\begin{equation}
	W_c^\dagger W_d = \frac{\gamma_c}{\pi} \mathds{1}_2 \delta_{cd}.
\end{equation} 
This means that there is no direct coupling between the different channels which is desirable since we are only interested in correlations that arise from the interaction zone itself. As has already been discussed in Refs. \cite{LSSS1995a,LSSS1995b} other choices for the average scattering matrix yield the same results for $M\ll N$. Contrary to the other two ensembles $(\beta=1,2)$ the product $W_c^\dagger W_d$ is not a scalar due to the quaternion structure.

Therefore, summarizing some basic properties of quaternions, for example found in \cite{ZHANG1997}, is called for. A quaternion $q$ can be represented by a complex $2 \times 2$ matrix of the following form
\begin{equation}\label{eqn:quaterniondef}
	q = \begin{bNiceMatrix}
		q_0 + i q_3 & q_2 + i q_1 \\
		-q_2 + i q_1 & q_0 - i q_3
	\end{bNiceMatrix} = q_0 \mathds{1}_2 + i \sum_{j=1}^{3} q_j \tau_j
\end{equation}
with the Pauli matrices $\tau_j$. We call $q$ a real quaternion if the coefficients $q_j, j=0,\ldots,3$ are real numbers. Furthermore, we see that the first and second column are not independent of each other, and it is possible to write
\begin{equation}\label{eqn:defY}
	q = \begin{bNiceMatrix}
		\widetilde{q} & Y \widetilde{q}^\star
	\end{bNiceMatrix}
	\quad \text{with} \quad \widetilde{q}=\begin{bNiceMatrix}
		q_0 + i q_3 \\
		-q_2 + i q_1
	\end{bNiceMatrix} \quad \text{and} \quad Y=\begin{bNiceMatrix}
	0 & -1 \\
	1 & 0
	\end{bNiceMatrix} .
\end{equation}
Where $Y (\cdot)^\star$ is the time reversal operation in half integer spin systems. We can identify the two components with a spin up $m=\uparrow$ and spin down $m=\downarrow$ contribution. Although, arbitrary we will stick to the convention that the first component is associated with $m=\uparrow$ and its time reversed partner corresponds to $m=\downarrow$.
Extending this notion to vectors or matrices of quaternions leaves us with two choices, either choosing the entries of the respective vector or matrix to be quaternions such that there is a $2\times 2$ substructure or constructing the vector or matrix with a $2 \times 2$ superstructure by inserting vectors or matrices for the coefficients in \cref{eqn:quaterniondef}. Both representation are linked by a simple permutation which orders the set of integers smaller or equal to $2M$ into even and odd numbers. Thus, we can choose the representation at will and for $H$ and $W_c$ we choose the latter such that we have
\begin{equation}
	H = \begin{bNiceMatrix}
		H_0 & H_1 \\
		-H_1^\star & H_0^\star
	\end{bNiceMatrix} \quad \text{and} \quad W_c = \begin{bNiceMatrix}
	\widetilde{W}_c & \ytens \widetilde{W}_c^\star
	\end{bNiceMatrix} ,
\end{equation}
where $H_0=H_0^\dagger$ is a Hermitian $N \times N$ matrix, $H_1=H_1^\trans$ is a complex skew-symmetric $N \times N$ matrix and $\widetilde{W}_c$ is a $2N$-dimensional complex vector. The Hamilton matrix $H$ is not only Hermitian, but it is also invariant under the transformation
\begin{equation}\label{eqn:selfdualprop}
	\ytens H^\star \ytenstrans = H .
\end{equation}
These are in total the properties of a real quaternion self-dual matrix. As this symmetry is also contained in $W_c W_c^\dagger$ we have that $G$ obeys \cref{eqn:selfdualprop} as well.
Inserting the $W_c$ we see that the off-diagonal $(a\neq b)$ elements $S_{ab}(E)$ are themselves quaternions and can be identified with the spin position of the incident and outgoing asymptotic waves
\begin{equation}
	\begin{bNiceMatrix}
		S_{a\uparrow b\uparrow} & S_{a\uparrow b\downarrow} \\
		S_{a\downarrow b\uparrow} & S_{a\downarrow b\downarrow}
	\end{bNiceMatrix}= -2\pi i\begin{bNiceMatrix}
		\widetilde{W}_a^\dagger G \widetilde{W}_b & \widetilde{W}_a^\dagger G \ytens \widetilde{W}_b^\star \\
		\widetilde{W}_a^\trans \ytenstrans G \widetilde{W}_b & \widetilde{W}_a^\trans \ytenstrans G \ytens \widetilde{W}_b^\star
	\end{bNiceMatrix} .
\end{equation}
We will calculate the distributions of the real and imaginary part of the four entries separately. 

We model the Hamiltonian of the interaction zone $H$ by a Gaussian random matrix from the Gaussian Symplectic Ensemble. For all three symmetry classes, we choose the corresponding distribution
\begin{equation}
	\mathcal{P}^{(\beta)}(H) \propto \exp\left(-\frac{\beta N}{4v^2} \text{Tr} H^2\right)
\end{equation}
with $N$ the aforementioned dimension of $H$, disregarding what number field the entries of $H$ are from, the parameter $v^2$ which can be chosen freely to match energy scales and  
\begin{equation}\label{eqn:Tr}
	\text{Tr} = \begin{cases}
		\tr & , \beta=1,2 \\
		\frac{1}{2} \tr & , \beta=4
	\end{cases} .
\end{equation}
The probability to find a matrix $H$ in the interval $[H,H+\dd H]$ is determined by $\mathcal{P}^{(4)}(H)\dd[H]$ with the volume element
\begin{equation}\label{eqn:volH}
	\dd[H] = \prod_{j\geq k} \dd H_{jk}^{(0)} \prod_{\alpha=1}^{\beta-1} \prod_{j>k} \dd H_{jk}^{(\alpha)} .
\end{equation}
Due to Kramers' degeneracy \cite{K1930} the additional factor $1/2$ occurs for $\beta=4$ in \cref{eqn:Tr} and is necessary to reproduce the correct Wigner semicircle law
\begin{equation}
	\rho(E) = \frac{1}{2\pi v^2} \sqrt{4v^2 - E^2} ,
\end{equation}
which is the eigenvalue density for $N\to\infty$.
\section{Distributions of Off-Diagonal Scattering Elements}
\label{sec:calcdistr}
We pursue an approach similar to the one taken in \cite{KNSGDMRS2013}. We introduce the real $(s=1)$ and imaginary $(s=2)$ part 
\begin{equation}
	\wp_s\left(S_{am bm\myprime}(E)\right) = \frac{1}{2i^{s-1}} \left(S_{am bm\myprime}(E) + (-1)^{s-1} S_{am bm\myprime}^\star(E) \right) 
\end{equation}
for $m,m\myprime\in\{\uparrow,\downarrow\}$.
The distribution is an ensemble average over a filter function,
\begin{equation}
	P_{s,m m\myprime}(x_s) = \int\dd[H] \mathcal{P}^{(4)}(H) \delta\left(x_s - \wp_s\left(S_{am bm\myprime}(E)\right)\right) ,
\end{equation}
where $\dd[H]$ is volume element defined in \cref{eqn:volH}. In the following we turn to the characteristic function since it allows for further analytical progress. The characteristic function contains the same information as the distribution itself and is readily available from experimental data as well. We have
\begin{equation}
	R_{s,m m\myprime}(k) = \int\dd[H] \mathcal{P}^{(4)}(H) \exp\left(-i k \wp_s\left(S_{am bm\myprime}(E)\right)\right)
\end{equation}
and observe that we can introduce the $4N$-dimensional vector $\widetilde{W}$
\begin{equation}
	\wp_s\left(S_{am bm\myprime}(E)\right) = \pi \widetilde{W}^\dagger A_s \widetilde{W} \quad \text{with} \quad A_s = \begin{bNiceMatrix}
		0 & (-i)^s G \\
		i^s G^\dagger & 0
	\end{bNiceMatrix}
\end{equation}
which is given by
\begin{equation}\label{eqn:wtilde}
	\widetilde{W} = \begin{cases}
		 \left(\widetilde{W}_a, \widetilde{W}_b\right)& , \, m=\uparrow, \, m\myprime=\uparrow \vspace{0.2cm}\\
		 \left(\widetilde{W}_a, \ytens \widetilde{W}_b^\star\right) & , \, m=\uparrow, \, m\myprime=\downarrow \vspace{0.2cm}\\
		 \left(\ytens \widetilde{W}_a^\star, \widetilde{W}_b\right) & , \, m=\downarrow, \, m\myprime=\uparrow \vspace{0.2cm}\\
		 \left(\ytens \widetilde{W}_a^\star, \ytens \widetilde{W}_b^\star\right) & , \, m=\downarrow, \, m\myprime=\downarrow 
	\end{cases} .
\end{equation}
We use the identity \cite{Berezin1987}
\begin{align}\label{eqn:gausscommanticommphase}
	&\exp\left(-i k \pi \widetilde{W}^\dagger A_s \widetilde{W}\right) \notag\\
	=& \int\dd[z] \int\dd[\zeta] \exp\left(\frac{i}{2} \left(z^\dagger \widetilde{W} + \widetilde{W}^\dagger z\right)\right) \exp\left(\frac{i}{4\pi k} \left(z^\dagger A_s^{-1} z + \zeta^\dagger A_s^{-1} \zeta\right)\right)
\end{align}
to rewrite our characteristic function in terms of Gaussian integrals over commuting and anticommuting degrees of freedom, here $z=(z_a, z_b)$ and $\zeta= (\zeta_a, \zeta_b)$ are $4N$-dimensional vectors containing commuting and anticommuting variables, respectively. This has the advantage that the exponent depends linearly on $H$ and the ensemble average over $H$ can be carried out. Before performing the ensemble average we write the integral more compactly by introducing the $8N$-dimensional supervector $\psi=(z,\zeta)$ and
\begin{equation}
	\widetilde{\bm{W}} = \left(\widetilde{W},0\right) \quad \text{and} \quad \bm{A}_s^{-1} = \mathds{1}_2 \otimes A_s^{-1}
\end{equation}
where the $0$ is understood as the $4N$-dimensional zero vector. With these definitions we can write
\begin{equation}\label{eqn:cfgaussian}
	R_{s,m m\myprime}(k) = \int\dd[\psi] \exp\left(\frac{i}{2}\left(\psi^\dagger \widetilde{\bm{W}} + \widetilde{\bm{W}}^\dagger \psi\right)\right) \int\dd[H] \mathcal{P}^{(4)}(H) \exp\left(\frac{i}{4\pi k} \psi^\dagger \bm{A}_s^{-1} \psi\right) .
\end{equation}
However, the inverse $\bm{A}_s^{-1}$ is still block off-diagonal
\begin{equation}
	A_s^{-1} = \begin{bNiceMatrix}
		0 & (-i)^s \left(G^\dagger\right)^{-1} \\
		i^s G^{-1} & 0
	\end{bNiceMatrix}
\end{equation}
which is undesirable for carrying out the ensemble average. Therefore, we employ the transformation
\begin{equation}
	z \to \Xi^+ z, \quad z^\dagger \to z^\dagger, \quad \zeta \to \Xi^- \zeta, \quad \zeta^\dagger \to \zeta\dagger
\end{equation}
with
\begin{equation}
	\Xi^{\pm} = \begin{bNiceMatrix}
		0 & \pm (-i)^s \mathds{1}_{2N} \\
		-i^s \mathds{1}_{2N} & 0
	\end{bNiceMatrix}
\end{equation}
which block diagonalizes $A_s^{-1}$, and we have
\begin{equation}
	\bm{A}_s^{-1} \left(\Xi^{+}\oplus\Xi^{-}\right) = \diag\left(-\left(G^{-1}\right)^\dagger, G^{-1}, - \left(G^{-1}\right)^\dagger, - G^{-1}\right) =: \bm{\Omega} 
\end{equation}
and
\begin{equation}
	\widetilde{\bm{W}}^\dagger \left(\Xi^{+}\oplus\Xi^{-}\right) =: \bm{\widetilde{U}}_s^\dagger .
\end{equation}
As the determinants of $\Xi^{+}$ and $\Xi^{-}$ are both unity, the Jacobian of this transformation is unity,
\begin{equation}\label{eqn:CFSupervector}
	R_s(k) = \int\dd[\psi] \exp\left(\frac{i}{2}\left(\psi^\dagger \widetilde{\bm{W}} + \widetilde{\bm{U}}_s^\dagger \psi\right)\right) \int\dd[H] \mathcal{P}^{(4)}(H) \exp\left(\frac{i}{4\pi k} \psi^\dagger \bm{\Omega} \psi\right).
\end{equation}
We separate the quadratic form $\psi^\dagger \bm{\Omega} \psi$ into an $H$-dependent part and $H$-independent part,
\begin{equation}
	\psi^\dagger \bm{\Omega} \psi = \tr H K + \psi^\dagger \bm{\Omega}_0 \psi 
\end{equation}
with
\begin{align}\label{eqn:defK}
	K =& z_a z_a^\dagger - z_b z_b^\dagger - \zeta_a \zeta_a^\dagger - \zeta_b \zeta_b^\dagger, \\
	\bm{\Omega}_0 =& E \, \diag(-1,1,-1,-1)\otimes\mathds{1}_{2N} + i \pi \, \diag(1,1,1,-1)\otimes \sum_{c=1}^{M} W_c W_c^\dagger.
\end{align}
Hence, the characteristic function takes the form
\begin{align}
	R_{s,m m\myprime}(k) =& \int\dd[\psi] \exp\left(\frac{i}{2}\left(\psi^\dagger \widetilde{\bm{W}} + \widetilde{\bm{U}}_s^\dagger \psi\right)\right) \exp\left(\frac{i}{4\pi k} \psi^\dagger \bm{\Omega}_0 \psi\right) \notag\\
	&\times \int\dd[H] \mathcal{P}^{(4)}(H) \exp\left(\frac{i}{4\pi k} \tr H K\right)
\end{align}
and we are left with the task to calculate the matrix Fourier transform of $\mathcal{P}^{(4)}(H)$.
\subsection{Hubbard-Stratonovich-Transformation}
The Fourier transform is non-trivial because $H$ and $K$ do not share the same symmetries. We solve this problem by incorporating the symmetries of $H$, see \cref{eqn:selfdualprop}, into $K$
\begin{equation}
	\begin{split}
	\tr H K =& \frac{1}{2} \left(\tr H K + \tr \ytens H^\trans \ytenstrans K \right) \\
	=& \frac{1}{2} \tr H \left(K + \ytenstrans K^\trans \ytens \right) =: \frac{1}{2} \tr H \widehat{K} .
	\end{split}
\end{equation}
It is easy to verify that indeed $\widehat{K}$ is not only Hermitian but also fulfils the property in \cref{eqn:selfdualprop}. The ensemble average is given by
\begin{equation}\label{eqn:ftordsuper}
	\int \dd[H] \mathcal{P}^{(4)}(H) \exp\left(\frac{i}{4\pi k}\tr H K\right) = \exp\left(-\frac{v^2}{2N (8\pi k)^2} \tr \widehat{K}^2\right)
\end{equation}
where the normalization of $P(H)$ exactly cancels the prefactor generated by the Fourier transform. We introduce
\begin{equation}\label{eqn:Amat}
	A = \begin{bNiceMatrix}
		z_a^\dagger \\
		z_a^\trans \ytenstrans \\
		z_b^\dagger, z_b^\trans \ytenstrans \\
		-\zeta_a^\dagger \\
		\zeta_a^\trans \ytenstrans \\
		-\zeta_b^\dagger \\
		\zeta_b^\trans \ytenstrans
	\end{bNiceMatrix} 
\end{equation} 
which is a rectangular supermatrix with zero Boson-Fermion and Fermion-Fermion blocks. We write $\widehat{K}$ as a function of the matrix $A$
\begin{equation}
	\widehat{K} = A^\dagger \widetilde{L} A \quad \text{with} \quad \widetilde{L} = \diag\left(1,-1,1,1\right)\otimes\mathds{1}_2 .
\end{equation}
We recall that $\widehat{K}$ itself is not a supermatrix but rather an ordinary $2N$-dimensional matrix with commuting and anticommuting entries. However, there exists a duality relation between ordinary and superspace \cite{Guhr2006}
\begin{equation}\label{eqn:dualityordsuperspace}
	\tr \widehat{K}^2 = \tr \left(A^\dagger \widetilde{L} A\right)^2 = \str \left(\widetilde{L}^{1/2} A A^\dagger \widetilde{L}^{1/2} \right)^2 = \str B^2 \quad \text{with} \quad B = \widetilde{L}^{1/2} A A^\dagger \widetilde{L}^{1/2} .
\end{equation}
Importantly, $B$ is a $4\vert 4 \times 4\vert 4$ supermatrix which means that we have drastically reduced the dimensionality of our problem using the above relation. 

The supermatrix $B$ contains the variables $z$ and $\zeta$ to second order which means in our expression they appear quartic. This is not desirable and calls for another transformation, which together with the Fourier transformation in \cref{eqn:ftordsuper} is often referred to as the Hubbard-Stratonovich-Transformation. We replace $B$ via a Fourier transformation in superspace by a supermatrix $\sigma$ with appropriate symmetries
\begin{equation}\label{eqn:hubstratrafo}
	\exp\left(-\frac{v^2}{2N (8\pi k)^2} \str B^2\right) = \mathcal{N}\int\dd[\sigma] \exp\left(-\frac{N(8\pi k)^2}{2v^2} \str \sigma^2\right) \exp\left(i \str\sigma B\right) 
\end{equation}
and some normalization $\mathcal{N}$.
We briefly discuss the symmetries of $B$ which result from the form in \cref{eqn:dualityordsuperspace}. Since the metric $\widetilde{L}$ is indefinite we see that $B^\dagger = \widetilde{L} B \widetilde{L}$. Additionally, we observe that an exchange symmetry exists, which originates from \cref{eqn:selfdualprop}, within the supermatrices $A$ and manifests as
\begin{equation}
	B = C B^\star C^\trans \quad \text{with} \quad C = \diag(Y,Y^\trans,X,X) \quad \text{and} \quad X = \begin{bNiceMatrix}
		0 & 1 \\
		1 & 0
	\end{bNiceMatrix} .
\end{equation} 
Additionally, since the left side of \cref{eqn:hubstratrafo} only depends on the supertrace of $B$ our model is invariant under transformations $T$ that preserve the symmetries of $B$. Such transformations $T$ have to fulfill
\begin{equation}\label{eqn:tcondition}
	T^\dagger \widetilde{L} T = \widetilde{L} \quad \text{and} \quad C T^\star C^\trans = T . 
\end{equation}
The collection of these transformations form the well known non-compact orthosymplectic supergroup $\text{UOSp}(2,2\vert 4)$. To implement the correct symmetries into $\sigma$ we let $\sigma= T^{-1} \sigma_D T$, where 
\begin{equation}
	\sigma_D = \mathfrak{V} \diag\left(\sigma_{\text{B},1}\mathds{1}_2, \sigma_{\text{B},2}\mathds{1}_2,i\sigma_{\text{F},1},i\sigma_{\text{F},2},i\sigma_{\text{F},3},i\sigma_{\text{F},4}\right) \mathfrak{V}^\dagger 
\end{equation}
with the real diagonal elements $\sigma_{\text{B},j}, \sigma_{\text{F},j}$ determined by the conditions in \cref{eqn:tcondition}. We notice the additional $i$ in front of the fermionic block introduced for convergence. We introduce $\mathfrak{V}=\diag\left(\mathds{1}_4,\mathds{1}_2 \otimes \mathfrak{v}\right)$ with
\begin{equation}
	\mathfrak{v} = \frac{1}{\sqrt{2}} \begin{bNiceMatrix}
		1 & i \\
		1 & -i
	\end{bNiceMatrix}
\end{equation}
which appears due to our choice to work with complex variables. It causes the fermionic block of $B$ to contain at first glance complex matrices which upon closer inspection are real matrices rotated into the complex plane by the transformation $V$. As $V$ is not a real transformation it can not simply be absorbed into the transformations $T$ and instead has to be accounted for separately. That is to say that by means of the transformations $T$ the fermionic diagonal blocks of $B$ can not be fully diagonalized as it would include an additional transformation which does not obey by the group structure of $T$. Therefore, we have to treat it separately. The underlying cause for this is that similar to the orthogonal case the symplectic case can be mapped to a real vector space, and it might be counterintuitive to not transform into a real basis. However, it turns out useful since it greatly simplifies the integrand. Alternatively, this transformation has to be applied to the vectors $\widetilde{W}$ resulting in four times as many terms. This is a particular feature that arises for the calculation of the distributions, and is not present for the correlations.
We will return to these symmetries and give more context at a later point. For now, we will use them to carry out the integrals over commuting and anticommuting degrees of freedom $z, \zeta$.
\subsection{Integration over Real Supervectors}
We take the rows of the supermatrix $A^\dagger$ and construct a vector
\begin{equation}
	\Psi = \left(z_a, \ytens z_a^\star, z_b, \ytens z_b^\star, \zeta_a, \ytens \zeta_a^\star, \zeta_b, \ytens \zeta_b^\star\right)
\end{equation}
to double the size of $\psi$. With this new vector we can write the supertrace over $\sigma$ and $B$ as a quadratic form
\begin{align}
	\str \sigma B =& \Psi^\dagger \left(\widetilde{L}^{1/2} \left(\tau^{(3)}\otimes\mathds{1}_4\right)\sigma^\trans \left(\tau^{(3)}\otimes\mathds{1}_4\right) \widetilde{L}^{1/2} \otimes\mathds{1}_{2N}\right) \Psi \notag\\
	=& \Psi^\dagger \left(\widetilde{L}^{1/2} \sigma \widetilde{L}^{1/2} \otimes\mathds{1}_{2N}\right) \Psi.
\end{align}
The second equality has to be understood in the sense that an additional substitution $\sigma\to\left(\tau^{(3)}\otimes\mathds{1}_4\right)\sigma^\trans \left(\tau^{(3)}\otimes\mathds{1}_4\right)$ in the integral is necessary. This substitution leaves the Gaussian-like term invariant and has a unit Berizinian. Additionally, we can write the $\widetilde{W}$ dependence more compactly
\begin{equation}
	\widetilde{\bm{U}}_s^\dagger \psi + \psi^\dagger \widetilde{\bm{W}}  = \widetilde{\bm{V}}_s^\trans \left(\mathds{1}_4 \otimes \diag\left(\mathds{1}_{2N}, \ytenstrans\right)\right) \Psi
\end{equation}
with
\begin{align}
	\widetilde{V}_{s,\uparrow\uparrow}^\trans =&
		\left[-i^s \widetilde{W}_b^\dagger \quad \widetilde{W}_a^\trans \quad (-i)^s \widetilde{W}_a^\dagger \quad \widetilde{W}_b^\trans\right] \notag\\
	\widetilde{V}_{s,\uparrow\downarrow}^\trans =&	\left[-i^s \widetilde{W}_b^\trans \ytenstrans \quad \widetilde{W}_a^\trans \quad (-i)^s \widetilde{W}_a^\dagger \quad \widetilde{W}_b^\dagger \ytenstrans\right] \notag\\
	\widetilde{V}_{s,\downarrow\uparrow}^\trans =&
	\left[-i^s \widetilde{W}_b^\dagger \quad \widetilde{W}_a^\dagger \ytenstrans \quad (-i)^s \widetilde{W}_a^\trans \ytenstrans \quad \widetilde{W}_b^\trans\right] \notag\\
	\widetilde{V}_{s,\downarrow\downarrow}^\trans =&
	\left[-i^s \widetilde{W}_b^\trans \ytenstrans \quad \widetilde{W}_a^\dagger \ytenstrans \quad (-i)^s \widetilde{W}_a^\trans \ytenstrans \quad \widetilde{W}_b^\dagger \ytenstrans\right]
\end{align}
and $\widetilde{\bm{V}}_s^\trans = [\widetilde{V}_s^\trans \quad 0]$. Similarly, we also use
\begin{equation}
	\frac{1}{2}\Psi^\dagger \bm{\mathcal{A}}_0^{-1} \Psi = \psi^\dagger \bm{\Omega}_0 \psi
\end{equation}
with 
\begin{equation}
	\bm{\mathcal{A}}_0^{-1} = E \, \left(\diag(-1,1,-1,-1)\otimes \mathds{1}_2\right)\otimes\mathds{1}_{2N} + i \pi \, \left(\diag(1,1,1,-1)\otimes\mathds{1}_2\right)\otimes \sum_{c=1}^{M} W_c W_c^\dagger
\end{equation}
which holds due to the symplectic nature of the matrices $W_c W_c^\dagger$. The supersymmetric formulation of our characteristic function is
\begin{align}\label{eqn:chrfctsupervector}
	R_{s,m m\myprime}(k) =& \ \mathcal{N} \int\dd[\sigma]  \exp\left(-\frac{N(8\pi k)^2}{2v^2} \str \sigma^2\right) \notag\\
	&\times\int\dd[\psi] \exp\left(\frac{i}{2} \widetilde{\bm{V}}_s^\trans \left(\mathds{1}_4 \otimes \diag\left(\mathds{1}_{2N}, \ytenstrans\right)\right) \Psi \right) \notag\\
	&\times \exp\left(i \Psi^\dagger \left(\widetilde{L}^{1/2} \sigma \widetilde{L}^{1/2} \otimes\mathds{1}_{2N} + \frac{1}{8\pi k} \bm{\mathcal{A}}_0^{-1}\right) \Psi\right) .
\end{align}
Although $\Psi$ does contain complex variables the entries appear in quaternion conjugated pairs which means that the integral is closely related to the Gaussian integral over a real vector. We use this fact by introducing a new supervector $\Phi = \left[x_a, y_a, x_b, y_b, \zeta_a, \zeta_a^\star, \zeta_b, \zeta_b^\star\right]$ where $x_c, y_c$ are $2N$-dimensional vectors and the scaled real, imaginary parts of $z_c$, respectively. The supervector $\Phi$ is related to $\Psi$ by $\Psi = D \Phi$ with
\begin{equation}
	 D= \left(\mathds{1}_4 \otimes \diag\left(\mathds{1}_{2N}, \ytens\right) \right) \diag\left(\mathfrak{v}\otimes\mathds{1}_{2N}, \mathfrak{v}\otimes\mathds{1}_{2N}, \mathds{1}_{8N} \right) .
\end{equation}
Incorporating this new set of variables we are left with a type of integral which was solved in \cite{NKSG2014}. We proceed similarly as in the prior work and outline the steps in \ref{app:intcommanticomm}. We arrive at the result
\begin{equation}\label{eqn:intcommanticommresult}
	R_{s,m m\myprime}(k) = \mathcal{N}\int\dd[\sigma]  \exp\left(-\frac{N(8\pi k)^2}{2v^2} \str \sigma^2\right) \sdet{}^{-1/2}\Sigma \exp\left(-\frac{i}{16} F_s^{(4)}\right)
\end{equation}
with the quantities
\begin{align}
	\Sigma =& \sigma_E \otimes \mathds{1}_{2N} + \frac{i}{8k} L \otimes W W^\dagger , \notag\\
	L =& \diag\left(1,-1,1,-1\right)\otimes \mathds{1}_2 , \notag\\
	\sigma_E =& \sigma - \frac{E}{8\pi k} \mathds{1}_8
\end{align}
and the phase
\begin{equation}
	F_s^{(4)} = \widehat{\bm{V}}_s^\trans \left(\widetilde{L}^{-1/2}\otimes\mathds{1}_{2N}\right) \Sigma^{-1} \left(\widetilde{L}^{-1/2}\otimes\mathds{1}_{2N}\right) \overline{\bm{V}}_s
\end{equation}
with the vectors 
\begin{align}
	\widehat{\bm{V}}_s^\trans =& \widetilde{\bm{V}}_s^\trans \left(\mathds{1}_4\otimes \diag\left(\mathds{1}_{2N},Y^\trans\otimes\mathds{1}_N\right)\right), \\
	\overline{\bm{V}}_s =& \left(\mathds{1}_4\otimes \diag\left(\mathds{1}_{2N},Y\otimes\mathds{1}_N\right)\right) \diag\left(X\otimes\mathds{1}_{2N},X\otimes\mathds{1}_{2N},\mathds{1}_{8N}\right) \widetilde{\bm{V}}_s .
\end{align}
We further simplify the expression 
\begin{equation}
	\sdet{}^{-1/2} \Sigma = \sdet{}^{-N} \sigma_E \prod_{c=1}^{M} \sdet{}^{-1}\left(\mathds{1}_8+\frac{i \gamma_c}{8\pi k} \sigma_E^{-1} L\right)
\end{equation}
and calculate the inverse of $\Sigma$
\begin{align}
	\Sigma^{-1} =& \sigma_E^{-1} \otimes \mathds{1}_{2N} - \sigma_E^{-1} \otimes \sum_{c=1}^{M} \frac{\pi}{\gamma_c} W_c W_c^\dagger + \sum_{c=1}^{M} \rho^{(c)} \otimes \frac{\pi}{\gamma_c} W_c W_c^\dagger \notag\\
	\quad \text{with} \quad \rho^{(c)} =& \left(\sigma_E + \frac{i\gamma_c}{8\pi k}L\right)^{-1}
\end{align}
which due to the nature of the vectors $\widehat{\bm{V}}_s$ and $\overline{\bm{V}}_s$ greatly simplifies the phase to,
\begin{equation}\label{eqn:fs4bf}
	F_s^{(4)} = i^{s+1} \frac{\gamma_b}{\pi} \left(\rho_{13}^{(b)} + \rho_{42}^{(b)}\right) + (-i)^{s+1} \frac{\gamma_a}{\pi} \left(\rho_{24}^{(a)} + \rho_{31}^{(a)}\right)
\end{equation}
for all $S_{am bm\myprime}$ in Boson-Fermion block notation. While this is only a small step we notice the importance of this result since it shows that the integrand is independent of $m$ and $m\myprime$. This means that all possible spin combinations are identically distributed. The result might be surprising at first but can be intuitively understood in a simple fashion. We assume our Hamilton matrix $H$ to be fully populated which means that both blocks $H_0$ and $H_0^\star$, corresponding to spin up and down parts of our system, respectively, are treated on equal footing. Furthermore, the coupling strength of an arbitrary channel $c$ is equal for both spin positions as $\lvert \widetilde{W}_c\rvert = \lvert \ytens \widetilde{W}_c^\star \rvert$. Hence, this means that by construction our model does not distinguish between the two spin orientations, and it is therefore gratifying to see that this is reflected in the distributions. In the following, we drop the indices $m, m\myprime$ as the characteristic function does not depend on them.

\subsection{Saddle Point Approximation}
\label{subsec:SaddlePointApprox}
At this point we return to our characteristic function. A saddle point approximation is possible to carry out the limit $N\to\infty$ since $N$ only appears as an explicit parameter in our integral. We closely follow the steps undertaken by Refs. \cite{VWZ1985} and \cite{NKSG2014} to carry out the saddle point approximation. We write
\begin{equation}\label{eqn:saddlepointint}
	R_s(k) = \mathcal{N}\int\dd[\sigma] \exp\left(-N \mathcal{L} - \delta\mathcal{L}\right)
\end{equation}
with the dominant part $\mathcal{L}$ which determines the saddle point
\begin{equation}
	\mathcal{L} = \frac{(8\pi k)^2}{2v^2} \str\sigma^2 + \str \ln\sigma_E
\end{equation}
and the perturbations
\begin{equation}
	\delta L = \sum_{c=1}^{M} \str \ln \left(\mathds{1}_8 + \frac{i\gamma_c}{8\pi k} \sigma_E^{-1} L\right) + \frac{i}{16\pi} \left(i^{s+1} \gamma_b \left(\rho_{13}^{(b)} + \rho_{42}^{(b)}\right) + (-i)^{s+1} \gamma_a \left(\rho_{24}^{(a)} + \rho_{31}^{(a)}\right)\right) .
\end{equation}
Therefore, the saddle point equation is given by the first variation of the dominant part
\begin{equation}\label{eqn:saddlepointeq}
	0 = 2\widetilde{r} \sigma + \sigma_E^{-1}
\end{equation}
with $\widetilde{r}=(8\pi k)^2/2v^2$. A diagonal solution is given by
\begin{equation}\label{eqn:diagonalsaddlepoint}
	\sigma_D^{0} = \frac{E}{16\pi k}\mathds{1}_8 + \frac{i \Delta}{16\pi k} L
\end{equation}
where $\Delta=\sqrt{4v^2-E^2}$ which reproduces the saddle points for $\beta=1,2$ up to a prefactor and Wigner's semicircle law \cite{KNSGDMRS2013}. This saddle point is not the only possible diagonal solution, and we justify our restriction to this specific solution in \ref{app:convergence}. 

However, these are surely not the only solutions. We know from the construction of $\sigma$ that we can apply transformations $T$ which leave the form invariant. We briefly mention here that the transformations $\mathfrak{V}$ do not appear here as they leave the diagonal solution of the saddle point equation unchanged due to $[\mathfrak{V},\sigma_D^0]=0$. That means we clearly have that $T^{-1} \sigma_D^0 T$ will also solve \cref{eqn:saddlepointeq}.

We take the same approach as Ref. \cite{VWZ1985} and inspect the generators $G$ of the transformations $T$ in $[1,2]$-block notation. There exists transformations generated by generators of the form $\diag\left(G_1,G_2\right)$ which commute with the saddle point solution \cref{eqn:diagonalsaddlepoint} as the solution is the identity up to a global sign in the respective blocks. We denote the transformations generated by the block-diagonal generators with $R$. They are themselves block-diagonal and form the group $\text{UOSp}(2\vert 2)\times \text{UOSp}(2\vert 2)$. As they leave the saddle point invariant the relevant parts of $T$ must be the parts up to the transformations $R$. These transformations $T_0$ are in the coset space $\text{UOSp}(2,2\vert 4)/\text{UOSp}(2\vert 2)\times \text{UOSp}(2\vert 2)$ and are generated by block-off-diagonal generators
\begin{equation}\label{eqn:GeneratorT0}
	T_0 = \exp\left(G_0\right) \quad \text{where} \quad G_0 = \begin{bNiceMatrix}
		0 & G_{0,12} \\
		G_{0,21} & 0
	\end{bNiceMatrix} . 
\end{equation}
These are the non-trivial transformations of the diagonal saddle point solution. Due to the coset structure and the generators we can reconstruct the full group by 
\begin{equation}
	T = R T_0 .
\end{equation}

We write $\sigma= T^{-1} \sigma_D T = T_0^{-1} P T_0$ where $P=R^{-1}\sigma_D R$ which means in the vicinity of the saddle point we have
\begin{equation}
	\sigma_D = \sigma_D^{0} + \delta\sigma_D^{0} \Rightarrow P = \sigma_D^{0} + R^{-1} \delta\sigma_D^{0} R = \sigma_D^{0} + \delta P
\end{equation}
since the diagonal solution to the saddle point equation $\sigma_D^{0}$ commutes with $R$ and also $V$. The deviation from the saddle point must get smaller with larger $N$ since the dominant part will yield quadratic dependence on $\delta P$ when expanded and therefore $\delta P$ has to go as $1/\sqrt{N}$. Therefore, it is useful to introduce $\delta P^\prime = \sqrt{N} \delta P$ to identify the non-vanishing terms in the exponent when carrying out the limit $N\to\infty$. 

We refer to 
\begin{equation}
	\sigma_G = T_0^{-1} \sigma_D^0 T_0 = \frac{E}{16\pi k} \mathds{1}_8 - \frac{\Delta}{16\pi k} Q, \quad Q = -i T_0^{-1} L T_0 
\end{equation}
as the "Goldstone" modes and $\delta P$ as the "massive" modes. Due to the choice of our parametrization the integral decouples when we expand our exponent in powers of $\delta P$. Therefore, we want to use the independent elements of $\sigma_G$ or rather $Q$ and $\delta P$ as our new variables of integration. The associated Berizinian is calculated in \ref{app:jacobgoldstoneheavy} and yields 
\begin{equation}
	\dd[\sigma] = \dd[\delta P] \dd\mu(Q) 
\end{equation} 
in the limit $N\to\infty$. The integration over the massive modes is trivial and amounts to a Gaussian integral in superspace which cancels a part of the normalization $\mathcal{N}$ and the remaining factor we will from now on denote $\alpha$. We are left with a non-linear sigma model of the form
\begin{equation}\label{eqn:nonlinearsigma}
	R_s(k) = \alpha\int\dd\mu(Q) \exp\left(-\frac{i}{16} F_s^{(4)}\right) \prod_{c=1}^{M} \sdet{}^{-1}\left(\mathds{1}_8 + \frac{i\gamma_c}{8\pi k}\sigma_{G,E}^{-1} L\right)
\end{equation}
which is structurally similar to the ones found for $\beta=1,2$, \textit{cf.} Ref. \cite{KNSGDMRS2013}. 

\subsection{Integration over the Saddle Point Manifold}
Up to this point it was sufficient to deal with the symmetries of $\sigma$ and an explicit parametrization was not necessary. However, at this point it becomes necessary. We derive a parametrization in \ref{app:parametrisation}. With the help of this parametrization we write 
\begin{equation}
	T_0 = \mathcal{U}^{-1} \begin{bNiceMatrix}
		\sqrt{\mathds{1}_4 - \sin^2\widehat{\theta}/2} & \sin\widehat{\theta}/2 \\
		-\sin\widehat{\theta}/2 & \sqrt{\mathds{1}_4 - \sin^2\widehat{\theta}/2}
	\end{bNiceMatrix} \mathcal{U}
\end{equation}
where we have introduced $\mathcal{U}=\diag\left(\widehat{U}^\dagger u_1, u_2\right)$ and
\begin{equation}
	\widehat{\theta} = \begin{bNiceMatrix}
		\widehat{\theta}_{\text{BB}} & 0 \\
		0 & \widehat{\theta}_{\text{FF}}
	\end{bNiceMatrix} \quad \text{with} \quad \widehat{\theta}_{\text{BB}} = i\theta\mathds{1}_2 \quad \text{and} \quad  \widehat{\theta}_{\text{FF}} = \begin{bNiceMatrix}
		\theta_1 & \theta_2 \\
		\theta_2 & \theta_1
	\end{bNiceMatrix} .
\end{equation}
Using these results we derive a block-wise diagonalization of $Q$ and arrive at 
\begin{equation}
	Q = -i \mathcal{U}^{-1} \begin{bNiceMatrix}
		\cos\widehat{\theta} & \sin\widehat{\theta} \\
		\sin\widehat{\theta} & -\cos\widehat{\theta}
	\end{bNiceMatrix} \mathcal{U} .
\end{equation}
With this parametrization we also determine the volume element $\dd\mu(Q)$ as the calculations are lengthy we defer to \ref{app:volumelemegoldstone} for a detailed discussion and only present the result here,
\begin{align} \label{eqn:volumegoldstone}
	\dd\mu(Q) =& \mathcal{B} \dd\theta\dd\theta_1\dd\theta_2\dd m \dd m_1 \dd m_1^\star \dd\phi_1 \dd\phi_2 \dd[\Upsilon],\notag\\
	\mathcal{B}=&2^3 \frac{\sin\theta_1\sin\theta_2}{\left(1+m^2+m_1m_1^\star\right)^3} \frac{\sinh^3\theta}{\left(\cos\left(\theta_1+\theta_2\right)-\cosh\theta\right)^2\left(\cos\left(\theta_1-\theta_2\right)-\cosh\theta\right)^2} ,
\end{align}
where $\dd[\Upsilon]$ contains all differentials of the anticommuting variables.

Furthermore, we use the parametrization to expand $\exp(-i F_s^{(4)}/16)$ in the anticommuting variables. Only terms which contain all anticommuting variables will yield a non-zero contribution to the integral. We present some intermediate steps of this calculation in \ref{app:phasecontribution}. Collecting these terms we arrive at
\begin{align}\label{eqn:fs4anticommintegration}
	\int\dd[\Upsilon] \exp\left(-\frac{i}{16} F_s^{(4)}\right) =& \frac{1}{(2\pi)^4} \Biggl[\frac{9}{16} k \partial_k + \frac{23}{16}k^2\partial_k^2 - \frac{1}{4}k^2 t_{ab} + \frac{5}{8}k^3\partial_k^3 - \frac{5}{16}k^3 t_{ab} \partial_k  \notag\\
	&+ \frac{1}{16} k^4\partial_k^4 - \frac{1}{16} k^4 t_{ab} \partial_k^2 + \frac{1}{16} k^4 t_{aa} t_{bb}\Biggr]\exp\left(-\frac{i}{16} k\kappa_0\right)
\end{align}
where we have introduced the quantities
\begin{align}
	\kappa_0 =& 8\sinh\theta \left[i^{s+1} \left(g_b^+ + \cosh\theta\right)^{-1} U_{11} + (-i)^{s+1} \left(g_a^+ +\cosh\theta\right)^{-1} U_{11}^\star\right] \notag\\
	t_{ab}=& \frac{\sin\left(\theta_1+\theta_2\right)}{g_a^+ + \cos\left(\theta_1+\theta_2\right)}  \frac{\sin\left(\theta_1+\theta_2\right)}{g_b^+ + \cos\left(\theta_1+\theta_2\right)} +  \frac{\sin\left(\theta_1-\theta_2\right)}{g_a^+ + \cos\left(\theta_1-\theta_2\right)} \frac{\sin\left(\theta_1-\theta_2\right)}{g_b^+ + \cos\left(\theta_1-\theta_2\right)} \notag\\
	t_{cc}=&\frac{\sin\left(\theta_1-\theta_2\right)\sin\left(\theta_1+\theta_2\right)}{\left(\cos\left(\theta_1-\theta_2\right)+g_c^+\right)\left(\cos\left(\theta_1+\theta_2\right)+g_c^+\right)}
\end{align}
with $g_c^+ = \left(\gamma_c^2 + v^2\right)/\left(\gamma_c \Delta\right)$. We point out that the quantities $t_{ab}, t_{cc}$ are closely related to the $t_c^{1}, t_c^2$ found for $\beta=2$ in \cite{KNSGDMRS2013} with $t_{ab}=t_a^1 t_b^1 + t_a^2 t_b^2$ and $t_{cc}=t_c^1 t_c^2$.
We can also further simplify the superdeterminant by using the saddle point equation and the explicit parametrization of $\sigma_G$ and arrive at
\begin{equation}
	\prod_{c=1}^{M} \sdet{}^{-1}\left(\mathds{1}_8 + \frac{i\gamma_c}{8\pi k}\sigma_{G,E}^{-1} L\right) = \prod_{c=1}^{M} \frac{\left(g_c^+ +\cos\left(\theta_1-\theta_2\right)\right)\left(g_c^+ +\cos\left(\theta_1+\theta_2\right)\right)}{\left(g_c^+ +\cosh\theta\right)^2}.
\end{equation}
Importantly, the integral no longer depends on the two orthogonal degrees of freedom $\phi_1, \phi_2$ and does not depend on one of the unitary degrees of freedom either. We are left with only five integrals. We solve the two remaining integrals over the unitary degrees of freedom. To that end it is useful to transform into a new set of variables
\begin{equation}
	U = \begin{bNiceMatrix}
		u e^{i \varphi_1} & \sqrt{1-u^2} e^{i\varphi_2} \\
		-\sqrt{1-u^2}e^{-i\varphi_2} & u e^{-i\varphi_1}
	\end{bNiceMatrix}
\end{equation}
which results in the following change of the volume element
\begin{equation}
	\frac{1}{\left(1+m^2+m_1 m_1^\star\right)^3} \dd m \dd m_1 \dd m_1^\star = \frac{1}{4} u \dd u \dd \varphi_1 \dd \varphi_2 .
\end{equation}
We arrive at
\begin{align}
	R_{s}(k) =& \frac{2\alpha}{2\pi} \int\dd[\widehat{\theta}] \frac{\sin\theta_1 \sin\theta_2 \sinh^3\theta}{\left(\cos\left(\theta_1-\theta_2\right)-\cosh\theta\right)^2 \left(\cos\left(\theta_1+\theta_2\right)-\cosh\theta\right)^2} \notag\\
	&\times \prod_{c=1}^{M} \frac{\left(g_c^+ +\cos\left(\theta_1-\theta_2\right)\right)\left(g_c^+ +\cos\left(\theta_1+\theta_2\right)\right)}{\left(g_c^+ +\cosh\theta\right)^2} \notag\\
	&\times \Biggl[\frac{9}{16} k \partial_k + \frac{23}{16}k^2\partial_k^2 - \frac{1}{4}k^2 t_{ab} + \frac{5}{8}k^3\partial_k^3 - \frac{5}{16}k^3 t_{ab} \partial_k + \frac{1}{16} k^4\partial_k^4 \notag\\
	&- \frac{1}{16} k^4 t_{ab} \partial_k^2 + \frac{1}{16} k^4 t_{aa} t_{bb}\Biggr]  \int_{0}^{1}\dd u \int_{0}^{2\pi} \dd\varphi_1 u \exp\left(-\frac{i}{16} \kappa_0\right) .
\end{align}
We solve the integral over $\varphi_1$ by expanding the exponential function in its power series and interchanging summation and integration which is permissible as the integral convergences uniformly. The integration over $\varphi_1$ becomes trivial, and we obtain the power series expression of the Bessel function of zeroth order. Then we apply the well known identity \cite{gradshteyn2007}
\begin{equation}
	\int\dd z z^{\nu+1} J_\nu(z) = z^{\nu+1} J_{\nu+1}(z)
\end{equation}
to carry out the integral over $u$. Thus, the characteristic function evaluates to
\begin{align}
	R(k) =& 2\alpha  \int\dd[\widehat{\theta}] \frac{\sin\theta_1 \sin\theta_2 \sinh^3\theta}{\left(\cos\left(\theta_1-\theta_2\right)-\cosh\theta\right)^2 \left(\cos\left(\theta_1+\theta_2\right)-\cosh\theta\right)^2}  \notag\\
	&\times\prod_{c=1}^{M} \frac{\left(g_c^+ +\cos\left(\theta_1-\theta_2\right)\right)\left(g_c^+ +\cos\left(\theta_1+\theta_2\right)\right)}{\left(g_c^+ +\cosh\theta\right)^2} \notag\\
	&\times \Biggl[\frac{9}{16} k \partial_k + \frac{23}{16}k^2\partial_k^2 - \frac{1}{4}k^2 t_{ab} + \frac{5}{8}k^3\partial_k^3 - \frac{5}{16}k^3 t_{ab} \partial_k + \frac{1}{16} k^4\partial_k^4 \notag\\
	&- \frac{1}{16} k^4 t_{ab} \partial_k^2 + \frac{1}{16} k^4 t_{aa} t_{bb}\Biggr] \frac{J_1(\omega_{ab} k)}{\omega_{ab} k}
\end{align}
with the quantity
\begin{equation}
	\omega_{ab} = \frac{\sinh\theta}{\sqrt{\left(g_a^+ + \cosh\theta\right)\left(g_b^+ + \cosh\theta\right)}} .
\end{equation}
We carry out the derivatives with help of the recurrence relations for Bessel functions and their derivatives, for example found in Ref. \cite{gradshteyn2007}, and arrive at the result
\begin{align} \label{eqn:rknoefetov}
	R(k) =& \frac{\alpha}{8}  \int\dd[\widehat{\theta}] \frac{\sin\theta_1 \sin\theta_2 \sinh^3\theta}{\left(\cos\left(\theta_1-\theta_2\right)-\cosh\theta\right)^2 \left(\cos\left(\theta_1+\theta_2\right)-\cosh\theta\right)^2}  \notag\\
	&\times\prod_{c=1}^{M} \frac{\left(g_c^+ +\cos\left(\theta_1-\theta_2\right)\right)\left(g_c^+ +\cos\left(\theta_1+\theta_2\right)\right)}{\left(g_c^+ +\cosh\theta\right)^2} \notag\\
	&\times k^2 \left(k^2\left(\omega_{ab}^4 + \omega_{ab}^2 t_{ab} + t_{aa} t_{bb}\right) \frac{J_1(\omega_{ab} k)}{\omega_{ab} k} - 2 \left(2\omega_{ab}^2 + t_{ab}\right) J_0(\omega_{ab} k)\right) .
\end{align}
From the definition of the characteristic function it must hold that $R(0)=1$ which is clearly not fulfilled by \cref{eqn:rknoefetov} as we have $R(0)=0$. This loss of normalization is due to the Efetov-Wegner terms arising from a change of coordinates in superspace. As seen from the singular value decomposition carried out in \ref{app:parametrisation} the singular values would contain additional even order terms of anticommuting variables. These additional contributions yield the aforementioned Efetov-Wegner terms. We discuss this in \ref{app:efetovwegner} using results obtained in Ref. \cite{VWZ1985} and find 
\begin{align}\label{eqn:finalresultmatelem}
	R(k) =& 1+\frac{1}{16}  \int_0^\infty\dd\theta \int_0^{\pi}\dd\theta_1 \int_{0}^{\pi/2}\dd\theta_2 \frac{\sin\theta_1 \sin\theta_2 \sinh^3\theta}{\left(\cos\left(\theta_1-\theta_2\right)-\cosh\theta\right)^2 \left(\cos\left(\theta_1+\theta_2\right)-\cosh\theta\right)^2}  \notag\\
	&\times\prod_{c=1}^{M} \frac{\left(g_c^+ +\cos\left(\theta_1-\theta_2\right)\right)\left(g_c^+ +\cos\left(\theta_1+\theta_2\right)\right)}{\left(g_c^+ +\cosh\theta\right)^2} \notag\\
	&\times k^2 \left(k^2\left(\omega_{ab}^4 + \omega_{ab}^2 t_{ab} + t_{aa} t_{bb}\right) \frac{J_1(\omega_{ab} k)}{\omega_{ab} k} - 2 \left(2\omega_{ab}^2 + t_{ab}\right) J_0(\omega_{ab} k)\right) .
\end{align}
The first term, \textit{i.e.} the unity, is the Efetov-Wegner term. The normalization $\alpha=1/2$ is fixed by the condition $R(0)=1$. The distributions follow according to
\begin{equation}
	P(x) = \frac{1}{2\pi} \int_{-\infty}^{\infty}\limits \dd k R(k) \exp\left(i k x\right) .
\end{equation}
This constitutes our final result. Some interesting observations follow. First, as was discussed above the characteristic function does not depend on the spin positions at all. Second, real and imaginary part are distributed equally. This is somewhat surprising as the saddle point manifolds of the GOE and GSE are closely related, and the distributions are not equal for the GOE. Also, the dependence on $s$ does not trivially drop out but rather is only revealed once the whole saddle point manifold is integrated besides the singular values. The intuition behind this difference is that we choose to inspect the elements of the quaternion $S_{ab}$ which hides its symplectic structure. If we would instead consider combinations of scattering matrix elements we suspect that the symplectic structure would give rise to GOE-like features. However, this is beyond the scope of the current analysis and work in progress. Third, while the integral might include poles at first glance when the denominator of the Berizinian vanishes this is not actually the fact since the rest of the integrand vanishes quicker. This can be seen by undertaking steps as outlined in \ref{app:efetovwegner}. At this point it is also possible to understand the necessity for the compactness of the fermionic block as otherwise the aforementioned zeroes of the nominator would not simply appear for isolated points and would render the integrals non-convergent as pointed out in Ref. \cite{wegner2016supermathematics}. Additionally, we find that the integral is invariant under the exchange of the channels $a$ and $b$. This is a consequence of the time reversal invariance of the system, which implies $S_{am bm\myprime} = S_{a\overline{m} b \overline{m}\myprime}$, where $\overline{m}$ is the spin orientations opposite to that of $m$. Since all spin orientations are distributed equally, this explains the invariance with respect to the channels.
\section{Distribution of Off-Diagonal Scattering Cross Sections}
\label{sec:scattercross}
Measurements of classical waves in swept-spectrum mode yield directly the scattering matrix elements, but in quantum experiments, only the cross sections $\sigma_{am bm\myprime}=\lvert S_{am bm\myprime} \rvert^2$ can be obtained in measurements. It is possible to obtain the distribution of the cross sections in a fashion similar to the characteristic function for the Gaussian Symplectic Ensemble. This has already been discovered for $\beta=1,2$ in \cite{KNSGDMRS2013}. The distribution of the scattering cross sections is given by
\begin{equation}
	p_{m m\myprime}(\sigma_{am bm\myprime}) = \int_{-\infty}^{+\infty}\limits\dd x_1 \int_{-\infty}^{+\infty}\limits\dd x_2 \delta(\sigma_{am bm\myprime}-x_1^2-x_2^2) P_{m m\myprime}(x_1,x_2)
\end{equation}
with the joint probability density function
\begin{equation}
	P_{m m\myprime}(x_1,x_2) = \int\dd[H] \mathcal{P}^{(4)}(H) \delta\left(x_1 - \text{Re} S_{am bm\myprime}\right) \delta\left(x_2 - \text{Im} S_{am bm\myprime}\right) .
\end{equation}
Following \cite{KNSGDMRS2013} we can formulate the distribution of cross sections in terms of the bivariate characteristic function $R(\bm{k})$, where $\bm{k}=k_1 + i k_2$ is a complex Fourier variable 
\begin{equation}\label{eqn:besseltransformcross}
	p_{m m\myprime}(\sigma_{am bm\myprime}) = \frac{1}{4\pi} \int\dd^2\bm{k} J_0\left(\sqrt{\sigma_{am bm\myprime}} \lvert \bm{k} \rvert\right) R_{m m\myprime}(\bm{k}) ,
\end{equation}
with the Bessel function of zeroth order $J_0(z)$. The bivariate characteristic function is given by
\begin{equation}
	R_{m m\myprime}(\bm{k}) = \int\dd[H] \mathcal{P}^{(4)}(H) \exp\left(-i\pi \widetilde{W}^\dagger A \widetilde{W}\right) \quad \text{with} \quad A = \begin{bNiceMatrix}
		0 & -i\bm{k}^\star G  \\
		i \bm{k} G^\dagger & 0
	\end{bNiceMatrix}
\end{equation}
where the $\widetilde{W}$ are the same as in \cref{eqn:wtilde}. We see that the advantage of this representation is the close relation to the univariate characteristic function which we have already calculated above. The only difference is that instead of a real $k$ with $i^s$ we have a complex $\bm{k}$ which appears in $A$ as $i \bm{k}$. Most of the calculations can be carried out analogously to the distribution of the scattering matrix elements. The aim of this section is to provide the necessary modifications.

Replacing the phase by Gaussian integrals over commuting and anticommuting variables we arrive, \textit{c.f.} \cref{eqn:cfgaussian}, at
\begin{equation}
	R_{m m\myprime}(\bm{k}) = \int\dd[\psi] \exp\left(\frac{i}{2}\left(\psi^\dagger \widetilde{\bm{W}} + \widetilde{\bm{W}}^\dagger \psi\right)\right) \int\dd[H] \mathcal{P}^{(4)}(H) \exp\left(\frac{i}{4\pi} \psi^\dagger \bm{A}^{-1} \psi\right) .
\end{equation}
We again want to employ some transformation such that the matrix
\begin{equation}
	A^{-1} = \begin{bNiceMatrix}
		0 & -i \left(\bm{k} G^\dagger\right)^{-1} \\
		i \left(\bm{k}^\star G\right)^{-1} & 0
	\end{bNiceMatrix}
\end{equation}
becomes block-diagonal. We employ
\begin{equation}
	z \to \Xi_z z \, , \quad z^\dagger \to z^\dagger \, , \quad \zeta \to \Xi_\zeta \zeta \, , \quad \zeta^\dagger \to \zeta^\dagger
\end{equation}
with the two matrices
\begin{equation}
	\Xi_z = \begin{bNiceMatrix}
		0 & -i\bm{k}^\star \\
		- i \bm{k} & 0
	\end{bNiceMatrix} \otimes\mathds{1}_{2N} \quad \text{and} \quad \Xi_\zeta = \begin{bNiceMatrix}
	0 & i\bm{k}^\star \\
	-i\bm{k} & 0
	\end{bNiceMatrix} \otimes \mathds{1}_{2N} 
\end{equation}
and recover the same matrix
\begin{equation}
	\bm{\Omega} = \bm{A}^{-1} \left(\Xi_z \oplus \Xi_\zeta\right) =  \diag\left(-\left(G^{-1}\right)^\dagger, G^{-1}, -\left(G^{-1}\right)^\dagger, - G^{-1}\right)
\end{equation}
as for the univariate characteristic function. The advantage of this procedure is that we shift the dependence on $\bm{k}$, as for the $s$ dependence in the univariate case, in the vectors $\widetilde{U}^\dagger = \widetilde{W}^\dagger \left(\Xi_z \oplus \Xi_\zeta\right)$. Both the parametrization of the saddle point manifold and also the corresponding volume element remain unchanged. The difference is within the phase which becomes for the bivariate characteristic function
\begin{equation}
	F^{(4)}(\bm{k}) = - \bm{k} \frac{\gamma_b}{\pi} \left(\rho_{13}^{(b)}+\rho_{42}^{(b)}\right) - \bm{k}^\star \frac{\gamma_a}{\pi} \left(\rho_{31}^{(a)}+\rho_{24}^{(a)}\right) .
\end{equation}
The remaining calculations are the same as for the univariate characteristic function, and we will omit them here. The resulting bivariate characteristic function is
\begin{align}\label{eqn:crosssec}
	R(\bm{k}) =& 1+\frac{1}{16}  \int_0^\infty\dd\theta \int_0^{\pi}\dd\theta_1 \int_{0}^{\pi/2}\dd\theta_2 \frac{\sin\theta_1 \sin\theta_2 \sinh^3\theta}{\left(\cos\left(\theta_1-\theta_2\right)-\cosh\theta\right)^2 \left(\cos\left(\theta_1+\theta_2\right)-\cosh\theta\right)^2}  \notag\\
	&\times\prod_{c=1}^{M} \frac{\left(g_c^+ +\cos\left(\theta_1-\theta_2\right)\right)\left(g_c^+ +\cos\left(\theta_1+\theta_2\right)\right)}{\left(g_c^+ +\cosh\theta\right)^2} \notag\\
	&\times \lvert \bm{k}\rvert^2 \left(\lvert \bm{k}\rvert^2\left(\omega_{ab}^4 + \omega_{ab}^2 t_{ab} + t_{aa} t_{bb}\right) \frac{J_1(\omega_{ab} \lvert \bm{k}\rvert)}{\omega_{ab} \lvert \bm{k}\rvert} - 2 \left(2\omega_{ab}^2 + t_{ab}\right) J_0(\omega_{ab} \lvert \bm{k}\rvert)\right) .
\end{align}
As in the case of the scattering matrix elements where there exists no dependence on $s$ we have here that the final result only depends on the absolute value of $\bm{k}$. Therefore, we can also deduct that the distribution of the phase of $\bm{k}$ is the uniform distribution over the unit circle. As expected from \cref{eqn:finalresultmatelem} as real and imaginary part are equally distributed. Additionally, a uniform distribution was obtained in Ref. \cite{KNSGDMRS2013} for unitary symmetry but not for orthogonal symmetry. In the latter case the bivariate characteristic function does not only depend on the absolute value $\vert \bm{k}\vert$ but also on the phase of $\bm{k}$. Also as for the scattering matrix elements we have dropped the indices $m$ and $m\myprime$ as the characteristic function is the same for all combinations. Finally, the calculation of the distribution from \cref{eqn:besseltransformcross} 
\begin{equation}
	p(\sigma_{ab}) = \frac{1}{2} \int_0^\infty\limits\dd\eta \, \eta J_0\left(\sqrt{\sigma_{ab}}\eta\right) R(\eta) .
\end{equation}
greatly simplifies using polar coordinates as the bivariate characteristic function only depends on the absolute value.

\section{Comparison with Numerical Data}
\label{sec:numerics}
\begin{figure}
	\subfloat{\includegraphics[scale=.8]{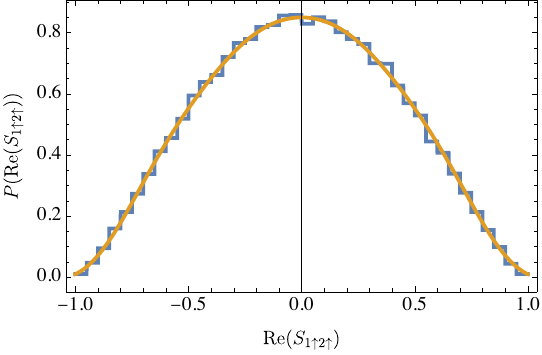}}
	\subfloat{\includegraphics[scale=.8]{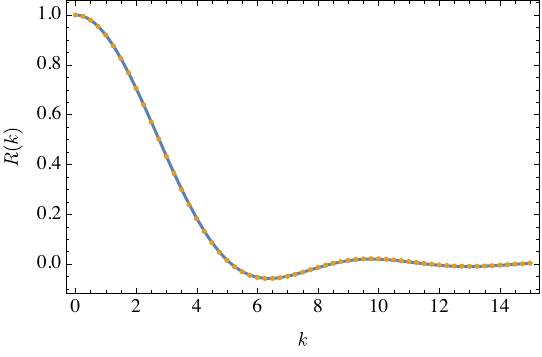}}
	\caption{PDF (left) and corresponding characteristic function (right) versus $\text{Re}\left(S_{1\uparrow 2\uparrow}\right)$ and k, respectively, with analytical results (orange), Monte-Carlo simulation (blue). The corresponding parameters are found in \cref{tab:parameters} (case I).}
	\label{fig:circ}
\end{figure}

\begin{figure}
	\subfloat{\includegraphics[scale=.8]{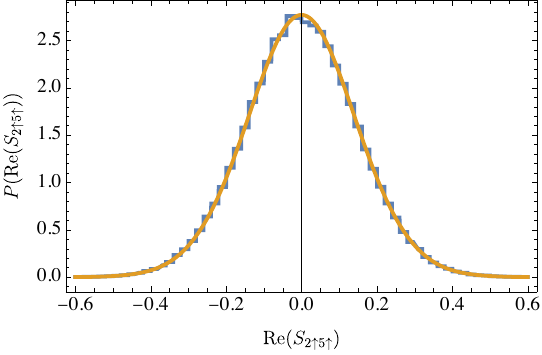}}
	\subfloat{\includegraphics[scale=.8]{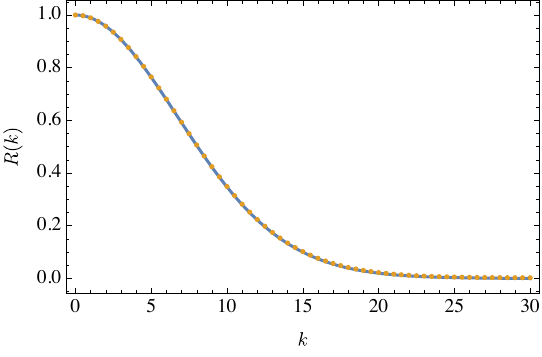}}
	\caption{PDF (left) and corresponding characteristic function (right) versus $\text{Re}\left(S_{2\uparrow 5\uparrow}\right)$ and k, respectively, with analytical results (orange), Monte-Carlo simulation (blue). The corresponding parameters are found in \cref{tab:parameters} (case II).}
	\label{fig:interm}
\end{figure}

\begin{figure}
	\subfloat{\includegraphics[scale=.8]{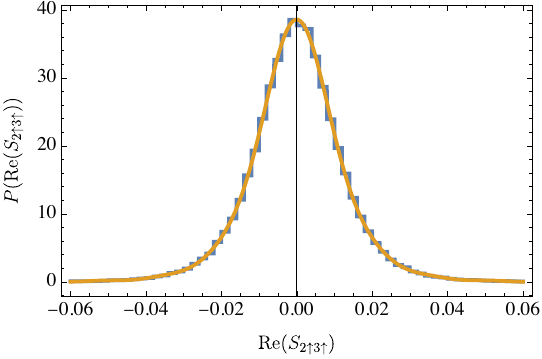}}
	\subfloat{\includegraphics[scale=.8]{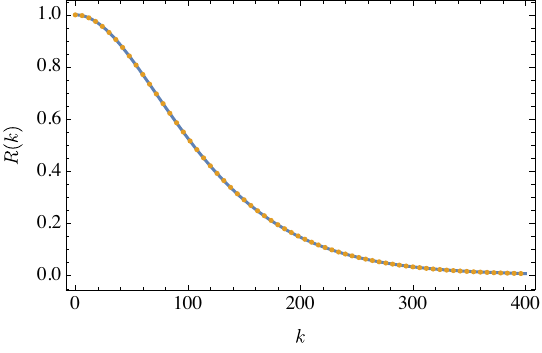}}
	\caption{PDF (left) and corresponding characteristic function (right) versus $\text{Re}\left(S_{2\uparrow 3\uparrow}\right)$ and k, respectively, with analytical results (orange), Monte-Carlo simulation (blue). The corresponding parameters are found in \cref{tab:parameters} (case III).}
	\label{fig:largeM}
\end{figure}

\begin{figure}
	\subfloat{\includegraphics[scale=.8]{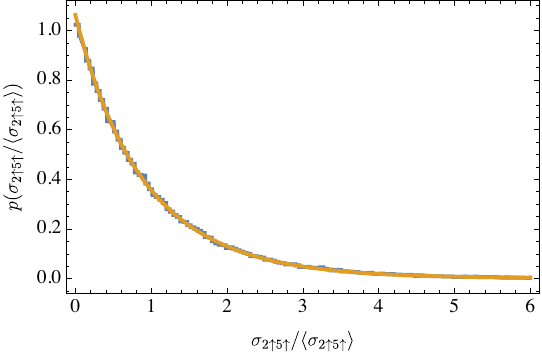}}
	\subfloat{\includegraphics[scale=.8]{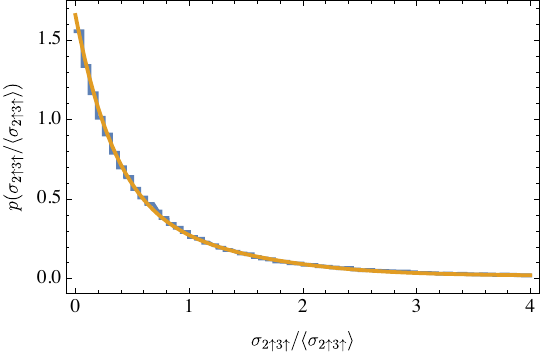}}
	\caption{Rescaled distribution of cross sections versus cross section rescaled to the mean with analytical results (orange), Monte-Carlo simulation (blue). The corresponding parameters are found in \cref{tab:parameters} (case II) and (case III) for the left and right figure, respectively.}
	\label{fig:cspdf}
\end{figure}

To validate our results we have performed Monte-Carlo simulations with $2N=600$ and a sample size of $2*10^5$ matrices using \textsc{Mathematica}. We compare the results of these simulations with numerical integration of our final results in \cref{eqn:finalresultmatelem,eqn:crosssec}. We perform these calculations for different set of parameters which are shown in \cref{fig:circ,fig:interm,fig:largeM,fig:cspdf}. We find full agreement in all cases for the distribution, characteristic function as well as the distribution of cross sections. As apparent from our choice of parameters, \textit{cf.} \cref{tab:parameters}, this is true for a wide range of channels with varying transmission and close or far away from the center of the semicircle. Additionally, as apparent from $p(0)\neq 1$ in \cref{fig:cspdf} we are not in the Ericson regime which characteristicly has $p(0) = 1$. Thus, our results remain valid in the regime of isolated resonances.

Comparisons with experimental setups and simulations of quantum graphs deserve a detached discussion as they include further technical details and are presented in a forthcoming publication \cite{GDG2025}.
\begin{table}
	\centering
	\caption{List of parameters used in Monte-Carlo simulations.}
	\begin{tabular}{c|c|c|c|c|c|c|c|c}
		case & $a$ & $m$ & $b$ & $m\myprime$ & $M$ & $T_c$ & $E$ & $v$ \\\hline
		I & $1$ & $\uparrow$ & $2$  & $\uparrow$ & $2$ & $T_1 = T_2 = 1$ & $0$ & $1$  \\\hline
		\multirow{2}{*}{II} & \multirow{2}{*}{$2$} & \multirow{2}{*}{$\uparrow$} & \multirow{2}{*}{$5$}  & \multirow{2}{*}{$\uparrow$} & \multirow{2}{*}{$6$} & $T_1=0.175, T_2=0.340, T_3=0.611,$ & \multirow{2}{*}{$1.2$} & \multirow{2}{*}{$1$}  \\
		& & & & & & $T_4=0.875, T_5=0.879, T_6= 0.780$ & &\\\hline
		III & $2$ & $\uparrow$ & $3$  & $\uparrow$ & $33$ & $T_c=0.039$ for $c=1,\ldots,32$ and $T_{33}=0.988$ & $0$ & $1$
	\end{tabular}
	\label{tab:parameters}
\end{table}

\section{Connection between Systems with Orthogonal and Symplectic Symmetry}
\label{sec:connectiongoegse}
While the final result for symplectic systems resembles those of the unitary system rather closely, there exists a striking resemblance for the saddle point manifolds with orthogonal systems. These similarities are already known from Ref. \cite{Zirnbauer1996} and also evident in the study of real forms of superalgebras \cite{Serganova1983}. While the results are not new, we aim to provide an intuitive understanding of the occurrence of the supergroup $\text{UOSp}(2,2\vert 4)$ for both orthogonal and symplectic symmetry. A more mathematical derivation of the two cases as fixed-point subgroups of the orthosymplectic supergroup $\text{OSp}(4\vert 4)$ is given in \ref{app:grouptheorygoegse}.

Let us start by recalling that the time reversal operator $\mathcal{T}$ in systems without or with integer spin is just the complex conjugation $\mathcal{K}$ whereas in system with half-integer spin it contains an additional rotation $Y$ in spin space, see \cref{eqn:defY},
\begin{equation}\label{eqn:timereversaloperator}
	\mathcal{T}^{(\beta=1)} = \mathcal{K} \quad \text{and} \quad \mathcal{T}^{(\beta=4)} = Y \mathcal{K} .
\end{equation}
Correspondingly, for $\beta=1$ the underlying number field is the real numbers whereas for $\beta=4$ it is the quaternions. As we are interested in the off-diagonal elements of the scattering matrix two sets of variables occur, those for the first channel $a$ and those for the second channel $b$. We obtain real and symplectic vectors of the form
\begin{equation}
	\bm{z}^{(\beta=1)} = \left(z, z^\star\right) \quad \text{and} \quad \bm{z}^{(\beta=4)} = \left(z, Y z^\star\right), \quad z = \left(z_a, z_b\right),
\end{equation} 
where we choose to rotate the real vector by an angle of $\pi/4$ into the complex plane. Heuristically, as the systems are invariant under time reversal this is reflected in the nature of the vectors as they not only contain $z$ but also their respective time reversed partner $\mathcal{T} z$. The full vector under time reversal exhibits
\begin{align}\label{eqn:timereversalcomm}
	\mathcal{T}^{(\beta=1)} \bm{z}^{(\beta =1)} =& \left(z^\star, z\right) = \left(\begin{bNiceMatrix}
		0 & 1 \\
		1 & 0
	\end{bNiceMatrix} \otimes\mathds{1}_2\right) \bm{z}^{(\beta =1)} \notag\\
	\mathcal{T}^{(\beta=4)} \bm{z}^{(\beta =4)} =& \left(Y z^\star, - z\right) = \left(\begin{bNiceMatrix}
		0 & 1 \\
		-1 & 0
	\end{bNiceMatrix} \otimes\mathds{1}_2\right) \bm{z}^{(\beta =4)}
\end{align}
with an additional negative sign for $\beta=4$ due to $\mathcal{T}^{(\beta=4)}\mathcal{T}^{(\beta=4)} = -\mathds{1}$ following from \cref{eqn:timereversaloperator}. Even though this as such seems to be of little relevance, there is an impact in superspace. Consider the anticommuting variables 
\begin{equation}
	\bm{\zeta}^{(\beta=1)} = \left(\zeta, \zeta^\star\right) \quad \text{and} \quad \bm{\zeta}^{(\beta=4)} = \left(\zeta, Y \zeta^\star\right), \quad \zeta = \left(\zeta_a, \zeta_b\right) .
\end{equation} 
Their behaviour under time reversal is very similar but with an important difference,
\begin{align}\label{eqn:timereversalanticomm}
	\mathcal{T}^{(\beta=1)} \bm{\zeta}^{(\beta =1)} =& \left(\zeta^\star, -\zeta\right) = \left(\begin{bNiceMatrix}
		0 & 1 \\
		-1 & 0
	\end{bNiceMatrix} \otimes\mathds{1}_2\right) \bm{\zeta}^{(\beta =1)},\notag\\
	\mathcal{T}^{(\beta=4)} \bm{\zeta}^{(\beta =4)} =& \left(Y \zeta^\star, \zeta\right) = \left(\begin{bNiceMatrix}
		0 & 1 \\
		1 & 0
	\end{bNiceMatrix} \otimes\mathds{1}_2\right) \bm{\zeta}^{(\beta =4)},
\end{align}
in the signs because $\left(\zeta^\star\right)^\star = - \zeta$. Following from \cref{eqn:timereversalcomm,eqn:timereversalanticomm} it is obvious that the roles of commuting and anticommuting variables swap with respect to the exchange symmetry for system with orthogonal and symplectic symmetry. Importantly, we did not arbitrarily choose these vectors because their elements crucially appear as the entries of $A$, \textit{cf.} \cref{eqn:Amat}. Thus, they determine the symmetries of $B$ which lead to the saddle point manifold. Extending the prior observation to the supermatrices we conclude that the bosonic and fermionic subspaces switch their respective roles. This explains why there is a striking resemblance to the orthogonal case, see, \textit{e.g.}, the diagonalisation of  $t_{12}$ in \ref{app:parametrisation}. 

Naturally, the question arises whether the two forms are identical, but this is not the case. On the one hand the subsets of supermatrices defined using the notion of the transposed in superspace are not invariant under the exchange of fermionic and bosonic subspaces. This is the case because the additional negative sign necessary in the definition of the transposed for supermatrices would occur in the Boson-Fermion block instead of the Fermion-Boson block after the exchange which clearly does not leave the space invariant. On the other hand and in this case much more importantly, the metric $\widetilde{L}$ defining the non-compactness does not depend on the symmetries of the system but rather is a consequence of convergence behaviour in superspace.

\section{Conclusions}
\label{sec:conclusion}
Using a new variant of the supersymmetry method, we calculated the distribution of scattering matrix elements and scattering cross sections in the case of symplectic symmetry exactly. This completes the derivation of these quantities for all Ensembles in Dyson's threefold way \cite{dyson1962}. Due to the symplectic structure, each scattering matrix element is itself a quaternion. Thus, we calculated eight different distributions, namely real and imaginary part for each possible spin combination. Interestingly, we found that the different spin orientations do not influence the distribution. Furthermore, real and imaginary part are equally distributed as for unitary symmetry, but in contrast to orthogonal symmetry. We explained both of these phenomena as features of our chosen model. The breaking of the spin symmetry is generally interesting is possible by breaking the symplectic features in the coupling, \textit{i.e.} different spin positions have different coupling strengths, or in the system itself by changing $H$. Both of these approaches go beyond the scope of the current investigation and are part of ongoing research. Additionally, our results enable the exact calculation of the distribution of scattering matrix elements in symplectic system in the Ericson regime, similar to the calculations performed in \cite{KDG2025} for unitary symmetry. We also provided a detailed approach to calculate the distributions of off-diagonal scattering matrix elements in symplectic systems and extensively discussed convergences issues and Efetov-Wegner terms. These analytic techniques extend previous works for the Gaussian Orthogonal and Gaussian Unitary Ensemble but to the best of our knowledge had not yet been formulated for the Gaussian Symplectic Ensemble. Additionally, we reformulated some of the approaches to such problems. Furthermore, we showed that our analytical result also hold up in a comparison with Monte-Carlo simulations. Finally, we provided intuitive and also rigorous understanding for the similarities between the saddle point manifolds of systems with symplectic and orthogonal symmetry following from the behaviour of bosonic and fermionic subspaces under time reversal.

\section*{Acknowledgments}
We thank A. Aldabag, B. Dietz, M. Kieburg and S. Köhnes for fruitful discussions. We also express our gratitude to the late Santosh Kumar for helpful comments during the early stages of this work and for providing \textsc{Mathematica} libraries.
This research was funded by the Deutsche Forschungsgemeinschaft (DFG, German Research Foundation) within the project Stochastic Quantum Scattering -- New Tools, New Aspects, DFG project number 540160740.
\newpage

\appendix
\section{Integral over commuting and anticommuting vectors}
\label{app:intcommanticomm}
To carry out the integral over the supervector $\psi$ we decouple the integral over commuting and anticommuting integral by exploiting symmetries of
\begin{equation}
	\widetilde{\Sigma} = D^\dagger \left(\widetilde{L}^{1/2} \otimes \mathds{1}_{2N}\right) \Sigma \left(\widetilde{L}^{1/2} \otimes \mathds{1}_{2N}\right) D
\end{equation}
which allow us to factor the integral
\begin{equation}
	\mathcal{I}_{\widetilde{\Sigma}} = \int\dd[\Phi] \exp\left(\frac{i}{2}\widetilde{\bm{V}}_s^\trans \diag\left(v\otimes\mathds{1}_{2N},v\otimes\mathds{1}_{2N},\mathds{1}_{8N}\right)\Phi\right) \exp\left(i\Phi^\trans J \widetilde{\Sigma}\Phi\right)
\end{equation}
into integrals over two Gaussian. This integral is the same as in \cref{eqn:chrfctsupervector} due to $\dd[\Phi]=\dd[\psi]$ following from the definition of $\Phi$ with an additional factor $1/\sqrt{2}$ and
\begin{equation}
	\Phi^\dagger = \Phi^\trans J \quad \text{with} \quad J =\diag\left(\mathds{1}_{8N}, J_2\right), \quad J_2 = \begin{bNiceMatrix}
		0 & -\mathds{1}_{2N} & 0 & 0 \\
		+\mathds{1}_{2N} & 0 & 0 & 0 \\
		0 & 0 & 0 & -\mathds{1}_{2N} \\
		0 & 0 & +\mathds{1}_{2N} & 0
	\end{bNiceMatrix} .
\end{equation}
We write out the quadratic form
\begin{equation}
	\Phi^\trans J \widetilde{\Sigma} \Phi = z^\trans \widetilde{\Sigma}_{11} z + z^\trans \widetilde{\Sigma}_{12}\zeta + \zeta^\trans J_2\widetilde{\Sigma}_{21} z + \zeta^\trans J_2\widetilde{\Sigma}_{22} \zeta
\end{equation}
with $z=[x_a,y_a,x_b,y_b]$ and $\zeta=[\zeta_a,\zeta_a^\star,\zeta_b,\zeta_b^\star]$ where we want to decouple the commuting and anticommuting degrees of freedom. Such a decoupling is made possible by the substitution
\begin{equation}
	\zeta \to \zeta - \widetilde{\Sigma}_{22}^{-1} \widetilde{\Sigma}_{21} z
\end{equation}
which also implies the analogous transformation for $\zeta^\trans$ since they are not independent. This substitution does not change the integral due to the invariance of integrals over anticommuting variables under a shift with another anticommuting variable. For this transformation to decouple the degrees of freedom as desired and solve the resulting integrals we require the symmetry properties
\begin{equation}
	\widetilde{\Sigma}_{11} = \widetilde{\Sigma}_{11}^\trans , \quad  J_2\widetilde{\Sigma}_{22} = - \left(J_2\widetilde{\Sigma}_{22}\right)^\trans \quad \text{and} \quad \widetilde{\Sigma}_{21}^\trans = - \widetilde{\Sigma}_{12} J_2 .
\end{equation}
We start with the first condition. It is clear that the contribution from $E\mathds{1}_8/8\pi k$ is symmetric since it is diagonal. Next we check the contribution of $WW^\dagger$ and see that its Boson-Boson-block has the form
\begin{align}
	& \left[D^\dagger \left(\widetilde{L}^{1/2}\otimes\mathds{1}_{2N}\right) L\otimes WW^\dagger \left(\widetilde{L}^{1/2}\otimes\mathds{1}_{2N}\right) D\right]_{11}\notag\\
	=&\mathds{1}_2\otimes \begin{bNiceMatrix}
		WW^\dagger + \ytenstrans W W^\dagger \ytens & i W W^\dagger - i \ytenstrans WW^\dagger \ytens \\
		-i W W^\dagger + i \ytenstrans WW^\dagger \ytens & W W^\dagger + \ytenstrans WW^\dagger \ytens
	\end{bNiceMatrix}
\end{align}
which is also symmetric due to the symplectic nature of the vectors $W$. We are left with investigating the symmetries of the contribution from $\sigma$. We recall that $\sigma^\dagger=\widetilde{L} \sigma \widetilde{L}$ and $\sigma=C \sigma^\star C^\trans$ which together gives $\sigma^\trans = C \widetilde{L} \sigma \widetilde{L} C^\trans$. We write
\begin{equation}
	D = \begin{bNiceMatrix}
		D_1 & 0 \\
		0 & D_2
	\end{bNiceMatrix} , \quad \widetilde{L} = \begin{bNiceMatrix}
	\widetilde{L}_1 & 0 \\
	0 & \widetilde{L}_2
	\end{bNiceMatrix} , \quad C = \begin{bNiceMatrix}
	C_1 & 0 \\
	0 & C_2
	\end{bNiceMatrix}
\end{equation}
and with that we have
\begin{equation}
	\left[D^\dagger \left(\widetilde{L}^{1/2}\otimes\mathds{1}_{2N}\right) \sigma\otimes\mathds{1}_{2N} \left(\widetilde{L}^{1/2}\otimes\mathds{1}_{2N}\right) D\right]_{11}^\trans = D_1^\trans \left(\widetilde{L}_1^{1/2}C_1\widetilde{L}_1 \sigma_{11} \widetilde{L}_1 C_1^\trans \widetilde{L}_1^{1/2}\right) \otimes \mathds{1}_{2N} D_1^\star
\end{equation}
which we can rewrite using that
\begin{align}
	D_1^\trans \left(\widetilde{L}_1^{1/2}C_1\widetilde{L}_1\right)\otimes\mathds{1}_{2N} =& D_1^\dagger \left(\widetilde{L}_2^{1/2}\otimes\mathds{1}_{2N}\right) \left(\mathds{1}_4\otimes\ytenstrans\right) \\
	\left(\widetilde{L}_1 C_1^\trans \widetilde{L}^{1/2}\right)\otimes\mathds{1}_{2N} D_1^\star =& \left(\mathds{1}_4\otimes\ytens\right) \left(\widetilde{L}_1^{1/2}\otimes\mathds{1}_{2N}\right) D_1
\end{align}
such that 
\begin{align}
	&\left[D^\dagger \left(\widetilde{L}^{1/2}\otimes\mathds{1}_{2N}\right) \sigma\otimes\mathds{1}_{2N} \left(\widetilde{L}^{1/2}\otimes\mathds{1}_{2N}\right) D\right]_{11}^\trans \notag\\
	=& \left[D^\dagger \left(\widetilde{L}^{1/2}\otimes\mathds{1}_{2N}\right) \sigma\otimes\mathds{1}_{2N} \left(\widetilde{L}^{1/2}\otimes\mathds{1}_{2N}\right) D\right]_{11}
\end{align}
and in total we see that we arrive at the desired $\widetilde{\Sigma}_{11}=\widetilde{\Sigma}_{11}^\trans$. Similarly, we can calculate the contributions to $\widetilde{\Sigma}_{22}$ where the energy dependent contributions are again trivial and the dependence on $W$ is given by
\begin{align}
	& J_2\left[D^\dagger \left(\widetilde{L}^{1/2}\otimes\mathds{1}_{2N}\right) L\otimes WW^\dagger \left(\widetilde{L}^{1/2}\otimes\mathds{1}_{2N}\right) D\right]_{22}\notag\\
	=&\mathds{1}_2\otimes \begin{bNiceMatrix}
		0 & -\ytenstrans WW^\dagger \ytens \\
		WW^\dagger & 0
	\end{bNiceMatrix} .
\end{align}
We turn to the $\sigma$ dependence and find
\begin{equation}
	\left[D^\dagger \left(\widetilde{L}^{1/2}\otimes\mathds{1}_{2N}\right) \sigma\otimes\mathds{1}_{2N} \left(\widetilde{L}^{1/2}\otimes\mathds{1}_{2N}\right) D\right]_{22}^\trans = D_2^\trans \left(\widetilde{L}_2^{1/2}C_2\widetilde{L}_2 \sigma_{22} \widetilde{L}_2 C_2^\trans \widetilde{L}_2^{1/2}\right) \otimes \mathds{1}_{2N} D_2^\star
\end{equation}
and with that we find
\begin{align}
	D_2^\trans \left(\widetilde{L}_2^{1/2}C_2\widetilde{L}_2\right)\otimes\mathds{1}_{2N} =& J_2 D_2^\dagger \left(\widetilde{L}_2^{1/2}\otimes\mathds{1}_{2N}\right) \left(\mathds{1}_4\otimes\ytenstrans\right) \\
	\left(\widetilde{L}_2 C_2^\trans \widetilde{L}_2^{1/2}\right)\otimes\mathds{1}_{2N} D_2^\star J_2^\trans =& \left(\mathds{1}_4\otimes -\ytens\right)\left(\widetilde{L}_2^{1/2}\otimes\mathds{1}_{2N}\right) D_2
\end{align}
such that
\begin{align}
	&\left[D^\dagger \left(\widetilde{L}^{1/2}\otimes\mathds{1}_{2N}\right) \sigma\otimes\mathds{1}_{2N} \left(\widetilde{L}^{1/2}\otimes\mathds{1}_{2N}\right) D\right]_{22}^\trans J_2^\trans \notag\\
	=& - J_2\left[D^\dagger \left(\widetilde{L}^{1/2}\otimes\mathds{1}_{2N}\right) \sigma\otimes\mathds{1}_{2N} \left(\widetilde{L}^{1/2}\otimes\mathds{1}_{2N}\right) D\right]_{22}
\end{align}
and we have $\left(J_2 \widetilde{\Sigma}_{22}\right)^\trans = - J_2\widetilde{\Sigma}_{22}$.

We are left with calculating the off-diagonal contributions, here neither $E$ nor $W$ appears since both do not have off-diagonal blocks. Thus, we turn to $\sigma$ again and see that
\begin{equation}
	\left[D^\dagger \left(\widetilde{L}^{1/2}\otimes\mathds{1}_{2N}\right) \sigma\otimes\mathds{1}_{2N} \left(\widetilde{L}^{1/2}\otimes\mathds{1}_{2N}\right) D\right]_{21}^\trans = D_1^\trans \left(\widetilde{L}_1^{1/2}C_1\widetilde{L}_1 \sigma_{12} \widetilde{L}_2 C_2^\trans \widetilde{L}_2^{1/2}\right) \otimes \mathds{1}_{2N} D_2^\star
\end{equation}
for which we use the same relations as before
\begin{align}
	D_1^\trans \left(\widetilde{L}_1^{1/2}C_1\widetilde{L}_1\right)\otimes\mathds{1}_{2N} =& D_1^\dagger \left(\widetilde{L}_2^{1/2}\otimes\mathds{1}_{2N}\right) \left(\mathds{1}_4\otimes\ytenstrans\right)\\
	\left(\widetilde{L}_2 C_2^\trans \widetilde{L}_2^{1/2}\right)\otimes\mathds{1}_{2N} D_2^\star =& \left(\mathds{1}_4\otimes -\ytens\right)\left(\widetilde{L}_2^{1/2}\otimes\mathds{1}_{2N}\right) D_2 J_2
\end{align}
and arrive at
\begin{align}
	&\left[D^\dagger \left(\widetilde{L}^{1/2}\otimes\mathds{1}_{2N}\right) \sigma\otimes\mathds{1}_{2N} \left(\widetilde{L}^{1/2}\otimes\mathds{1}_{2N}\right) D\right]_{21}^\trans \notag\\
	=& - \left[D^\dagger \left(\widetilde{L}^{1/2}\otimes\mathds{1}_{2N}\right) \sigma\otimes\mathds{1}_{2N} \left(\widetilde{L}^{1/2}\otimes\mathds{1}_{2N}\right) D\right]_{12} J_2
\end{align}
such that we have $\widetilde{\Sigma}_{21}^\trans = - \widetilde{\Sigma}_{12} J_2$ which are the desired relations, and we have shown that our $\widetilde{\Sigma}$ has the correct symmetries to apply the above relations to our integral.

We have shown that $\widetilde{\Sigma}$ has the desired symmetries and decouple the degrees of freedom which results in
\begin{equation}
	\Phi^\trans J \widetilde{\Sigma} \Phi = z^\trans \left(\widetilde{\Sigma}_{11}-\widetilde{\Sigma}_{12}\widetilde{\Sigma}_{22}^{-1}\widetilde{\Sigma}_{21} \right) z + \zeta^\trans J_2 \widetilde{\Sigma}_{22} \zeta .
\end{equation}
We apply two identities \cite{wegner2016supermathematics}
\begin{equation}
	\int\dd[z] \exp\left(i z^\trans a z + i b^\trans z\right) = \sqrt{\det\left(\frac{i \pi}{a}\right)} \exp\left(-\frac{i}{4} b^\trans a^{-1} b\right)
\end{equation}
where $a=a^\trans$ is a symmetric matrix $z, b$ real vectors and
\begin{equation}
	\int\dd[\zeta] \exp\left(i\zeta^\trans a \zeta\right) = \sqrt{\det\left(\frac{a}{i\pi}\right)}
\end{equation}
where $a=-a^\trans$ is a skew-symmetric matrix. Furthermore, we see that our required symmetry relations are also necessary to be able to apply these integral identities. We then have
\begin{align}
	\mathcal{I}_{\widetilde{\Sigma}} =& \det\left(\frac{i\pi}{\widetilde{\Sigma}_{11}-\widetilde{\Sigma}_{12}\widetilde{\Sigma}_{22}^{-1}\widetilde{\Sigma}_{21} }\right)^{1/2} \det\left(\frac{\widetilde{\Sigma}_{22}}{i\pi}\right)^{1/2}\notag\\
	&\times\exp\left(\frac{i}{16}\widetilde{V}_s^\trans \diag\left(v\otimes\mathds{1}_{2N},v\otimes\mathds{1}_{2N}\right)\left[\widetilde{\Sigma}^{-1}\right]_{11}\diag\left(v^\trans\otimes\mathds{1}_{2N},v^\trans\otimes\mathds{1}_{2N}\right) \widetilde{V}_s\right) .
\end{align}
We identified
\begin{equation}
	\left(\widetilde{\Sigma}_{11}-\widetilde{\Sigma}_{12}\widetilde{\Sigma}_{22}^{-1}\widetilde{\Sigma}_{21}\right)^{-1} = \left[\widetilde{\Sigma}^{-1}\right]_{11}
\end{equation}
such that we have
\begin{align}
	&\widetilde{V}_s^\trans \diag\left(v\otimes\mathds{1}_{2N},v\otimes\mathds{1}_{2N}\right)\left(\widetilde{\Sigma}_{11}-\widetilde{\Sigma}_{12}\widetilde{\Sigma}_{22}^{-1}\widetilde{\Sigma}_{21}\right)^{-1}\diag\left(v^\trans\otimes\mathds{1}_{2N},v^\trans\otimes\mathds{1}_{2N}\right) \widetilde{V}_s \notag\\
	=& \widehat{\bm{V}}_s^\trans \left(\widetilde{L}^{1/2}\otimes\mathds{1}_{2N}\right) \Sigma^{-1} \left(\widetilde{L}^{1/2}\otimes\mathds{1}_{2N}\right) \overline{\bm{V}}_s
\end{align} 
and
\begin{equation}
	\det\left(\frac{\widetilde{\Sigma}_{22}}{\widetilde{\Sigma}_{11}-\widetilde{\Sigma}_{12}\widetilde{\Sigma}_{22}^{-1}\widetilde{\Sigma}_{21}}\right)^{1/2} = \sdet\Sigma^{-1/2}
\end{equation}
which together results in the claimed \cref{eqn:intcommanticommresult}.

\section{Calculation of the Berizinian for Goldstone and heavy modes}
\label{app:jacobgoldstoneheavy}
In the saddle point approximation, we find that the integrand decouples with respect to the Goldstone and massive modes. In order for the full integral to decouple, we need to show that the Berezinian does as well. To that end, it is necessary to calculate the transformation of the volume element $\dd[\sigma]$ that comes with letting $\sigma = T_0^{-1} P T_0$, where $P=\sigma_D^{0} + \delta P$. We calculate the associated Berezinian by means of the elementary length element \cite{Hua1963,Efetov1997}. The matrix differential is
\begin{equation}
	\dd \sigma = T_0^{-1} \left(\dd P - \left[\dd T_0 \, T_0^{-1},P\right]\right)T_0,
\end{equation}
which gives the elementary length element
\begin{equation}
	\str \dd\sigma^2 = \str \dd P^2 + \str\left[\dd T_0 \, T_0^{-1} , P\right]^2 - 2 \str \dd P \left[\dd T_0 \, T_0^{-1}, P\right] .
\end{equation} 
In the limit $N\to\infty$, we have
\begin{equation}
	\lim\limits_{N\to\infty} P = \sigma_D^0 ,
\end{equation}
as $\delta P = \delta P\myprime/\sqrt{N}$, where $\delta P\myprime$ is independent of $N$. Thus, the elementary length element is
\begin{equation}\label{eqn:limitElemntaryLengthElement}
	\lim\limits_{N\to\infty}\str \dd\sigma^2 = \str \dd\delta P^2 + \str\left[\dd T_0 \, T_0^{-1},\sigma_D^{0}\right]^2 - 2\str \dd\delta P\left[\dd T_0 \, T_0^{-1} , \sigma_D^{0}\right] .
\end{equation}
Importantly, $\dd\delta P$ does not vanish in the limit $N\to\infty$ as the transformation from $\dd\delta P$ to $\dd\delta P^\prime$ has unit Berezinian. The last term in \cref{eqn:limitElemntaryLengthElement} vanishes since $\delta P$ is block diagonal and so is $\dd\delta P$ and $\sigma_D^{0}$ is the unit matrix in the respective blocks. Thus, $\dd\delta P$ and $\sigma_D^0$ commute, and we have due to the cyclic invariance
\begin{equation}
	\str \dd\delta P\left[\dd T_0 \, T_0^{-1} , \sigma_D^{0}\right] = \str \dd T_0 \, T_0^{-1} \left[\dd\delta P, \sigma_D^0\right] = 0
\end{equation}
and the supertrace vanishes, as claimed above. Inserting $\sigma_D^{0}$ leaves us with
\begin{equation}
	\str\dd\sigma^2 = \str \dd\delta P^2 + \frac{\Delta}{16\pi k} \str \dd Q^2 ,
\end{equation} 
which means the volume element in the limit $N\to\infty$ is given by
\begin{equation}
	\dd[\sigma] = \dd[\delta P] \dd\mu(Q) .
\end{equation}

\section{Convergence and saddle point}
\label{app:convergence}
We recall the identity \EQref{eqn:gausscommanticommphase}, used to replace a phase by a Gaussian integral over commuting variables, 
\begin{align}
	&\exp\left(-ik\pi \widetilde{W}^\dagger A_s \widetilde{W}\right) \notag\\
	=& \det\left(-ik\pi A_s\right) \int\dd[z] \exp\left(\frac{i}{2} z^\dagger \widetilde{W} + \widetilde{W}^\dagger z\right) \exp\left(\frac{i}{4\pi k} z^\dagger A_s^{-1} z\right) .
\end{align}
This identity is only valid if the matrix 
\begin{align}
	A_s =& \begin{bNiceMatrix}
		0 & (-i)^s G \\
		i^s G^\dagger & 0
	\end{bNiceMatrix} , \notag\\
	G =& E \mathds{1}_{2N} - H + i\pi \sum_{c=1}^{M} W_c W_c^\dagger
\end{align}
is non-singular. Unfortunately, this is not the case, as we integrate over $H$ and some of the eigenvalues eventually coincide with the energy $E$. This problem is not resolved by the additional term $i W W^\dagger$, as this matrix is of rank $2M$, which we, by construction, assume to be much smaller than our system size $2N$. Therefore, we have to shift the energy into the complex plane to avoid the resonances of the system and make the integrals convergent. We do this by replacing 
\begin{equation}\label{eqn:contourgaussian}
	E \mathds{1}_4 \otimes \mathds{1}_{2N} \to E \mathds{1}_4 \otimes \mathds{1}_{2N} - i \eta L \otimes \mathds{1}_{N}
\end{equation}
in \cref{eqn:CFSupervector} or alternatively adding an appropriately modified infinitesimal shift into the complex axis to the diagonals of $A_s$. 
Technically it is not necessary to do this for the anticommuting variables, since no convergence issues can arise from their integration, but we want to preserve the symmetry between bosonic and fermionic subspaces. Thus, we have fixed a contour in the complex plane, not only for $H$ but also as a consequence for $\sigma$. 

In the scope of our considerations, we have not yet addressed the sign ambiguity in the transformations $\Xi^{\pm}$, which then determine the signs of $K$ in \cref{eqn:defK}. Furthermore, we did not justify the explicit choice of the saddle point $\sigma_D^0$. These two questions are inherently linked, which we show in the following. We start by discussing the sign ambiguity in \cref{sec:SignAmbiguity} and then justify the choice of saddle point in \cref{sec:CorrectSaddlePoint}.

Here, we aim to extend the formalism derived in Ref. \cite{VWZ1985} to the \emph{GSE} and resolve the sign ambiguity and convergence questions. The content of this appendix already exist in a much more general and rigorous formulation in Ref. \cite{Zirnbauer1996}, which deals with all ten symmetry classes. However, we provide a more intuitive and specific approach, which we think is not only necessary but also helpful in order to fully understand the formalism derived in this work. Thus, while the results are not new, we still decide to give a detailed discussion.

\subsection{Resolving the Sign Ambiguity of the Metric}
\label{sec:SignAmbiguity}
Manipulating the sign of the entries of $\Xi^\pm$, it is possible to implement different realisations of the metric $\widetilde{L}$, which determines the transformations $T$. It is natural to ask why our choice should be the correct one. The reason for this is multifaceted and has been discussed in Ref. \cite{VWZ1985} for \emph{GOE} systems. We show that this discussion extends to the \emph{GSE} and reformulate the arguments in a more compact way. 

Let us return to the matrix
\begin{equation}
	K = z_a z_a^\dagger - z_b z_b^\dagger - \zeta_a \zeta_a^\dagger - \zeta_b \zeta_b^\dagger,
\end{equation} 
defined in \cref{eqn:defK}. We notice that transformations that exchange the commuting variables $z_a, z_b$ have to adhere to an indefinite metric, \textit{i.e.} non-compact transformations, as the first two terms of $K$ have opposite signs. Transformations that exchange the anticommuting variables $\zeta_a, \zeta_b$ are unitary transformation, since both terms come with the same sign. Hence, we find that the relative signs are determined by the signs of $\Xi^\pm$. 

Our choice to construct $\Xi^\pm$ in such a way that the exchange of commuting variables involves non-compact transformations is motivated by the results of Ref. \cite{SW1980}, in which the authors showed that for commuting variables only an indefinite metric results in a non-trivial non-linear sigma model. As evident from Ref. \cite{VWZ1985}, this also extends to chaotic scattering problems, and we adopt this non-compactness as well. We give a more intuitive explanation at the end of this chapter and show that a compact formulation only yields trivial results.

Hence, we are left with justifying the compactness of the transformations exchanging anticommuting variables in $K$. Before we are capable of a proper justification, we need to address the problem of convergence of the integration in \cref{eqn:hubstratrafo}. We know that implementing the symmetries of $K$ into $\sigma= T^{-1} \sigma_D T$ involves the non-compact transformations $T\in\text{UOSp}(2,2\vert 4)$. Thus, the integral
\begin{equation}\label{eqn:FTSuperSpaceConvergence}
	\exp\left(- \frac{v^2}{2N \left(8 \pi k\right)^2} \str B^2\right) = \mathcal{N} \int\dd[\sigma] \exp\left(- \frac{N (8\pi k)^2}{2v^2} \str\sigma^2\right) \exp\left(i \str \sigma B\right)
\end{equation}
does not converge. To illustrate this point, we recall that the supertrace is given by
\begin{equation}
	\str \begin{bNiceMatrix}
		\sigma_{\text{BB}} & \sigma_{\text{BF}} \\
		\sigma_{\text{FB}} & i\sigma_{\text{FF}}
	\end{bNiceMatrix}^2 = \tr \sigma_{\text{BB}}^2 + 2 \tr \sigma_{\text{BF}} \sigma_{\text{FB}} + \tr \sigma_{\text{FF}}^2 .
\end{equation} 
We know that the Boson-Fermion and Fermion-Boson blocks never cause any convergence issues, and we also neglect the Fermion-Fermion block for now. The trace of the square of the Boson-Boson block only depends on the diagonal elements of $\sigma_{\text{BB}}$. Meaning the Gaussian factor in \cref{eqn:FTSuperSpaceConvergence} assuredly renders the integrals over the diagonal elements $\sigma_D$ convergent. However, since $\sigma$ is diagonalised by non-compact transformations $T$, their integrals do not converge. This is not helped by the factor $\exp\left(i \str \sigma B\right)$. Consider $\sigma=T^{-1}\sigma_D T$, and absorb the transformations $T$ into $B$, as they are constructed in such a way that they leave the form of $B$ invariant. By construction the diagonal elements of $\sigma_D$ and $B$ are real. Thus, the supertrace only depends on the diagonal elements of $\sigma_D$ and $B$, and $\exp\left(i\str\sigma B\right)$ amounts to a phase. Hence, it does not contribute toward making the integration over $T$ convergent. 

Therefore, we are forced to implement an additional factor that makes our integral convergent. This factor is
\begin{equation}\label{eqn:convfac}
	\exp\left(i \alpha \frac{N}{4} \str \sigma \widetilde{L}_{\text{BB}}\right),
\end{equation}
with $\widetilde{L}_{\text{BB}}=\diag\left(\mathds{1}_2,-\mathds{1}_2,0_4\right)$, the projection of the metric $\widetilde{L}$ onto the Boson-Boson block. Inserting the parametrisation derived in \cref{app:parametrisation}, we get in $[1,2]$ block notation
\begin{align}\label{eqn:fullconvexp}
	&\exp\Biggl(i \alpha \frac{N}{4} \Biggl\{\tr \left[P_{11}(\mathds{1}_4 + 2 t_{12} t_{21})\right]_{\text{BB}} \notag\\
	&\bralign{\exp\Biggl(i \alpha \frac{N}{2}}- \tr \left[P_{22} (\mathds{1}_4 + 2 t_{21} t_{12})\left(\sigma^{(3)} \otimes \mathds{1}_2\right)\right]_{\text{BB}}\Biggr\}\Biggr) .
\end{align}
This factor by itself does not guarantee convergence, but shifting $P$ by an infinitesimal shift $\gamma$ into the complex plane
\begin{equation}\label{eqn:convtrafobb}
	P_{\text{BB},11} \to P_{\text{BB},11} + i \gamma \mathds{1}_2 \quad \text{and} \quad P_{\text{BB},22} \to P_{\text{BB},22} - i \gamma \mathds{1}_2 
\end{equation}
results in the additional factor
\begin{equation}\label{eqn:convgenfac}
	\exp\left(-\alpha\gamma \frac{N}{4} \tr\left[\mathds{1}_4+2t_{12}t_{21} + \mathds{1}_4 + 2 t_{21} t_{12}\right]_{\text{BB}}\right).
\end{equation}
We omitted $\sigma^{(3)}\otimes \mathds{1}_2$, since its Boson-Boson block is the unit matrix. Conveniently, \cref{eqn:convgenfac} only depends on the singular values of $t_{12}$, \textit{cf.} \cref{eqn:DiagonalizationT12}, such that we find that the exponent of
\begin{equation}
	\exp\left(-\alpha\gamma \frac{N}{4} \tr\left[\mathds{1}_4+2t_{12}t_{21} + \mathds{1}_4 + 2 t_{21} t_{12}\right]_{\text{BB}}\right) = \exp\left(-\alpha\gamma N \cosh\theta\right)
\end{equation}
is always negative, thus, rendering the integrals over the non-compact singular value $\theta$ of $t_{12}$ convergent. Performing the limit $\gamma\to 0$ and $\widetilde{\alpha}=\alpha N\to 0$, after carrying out the integration over the saddle point manifold, leads to the final result in \cref{eqn:finalresultmatelem}. This limit is trivial, as the integrals over the singular values are already convergent and the convergence generating factor simply drops out. This explains why we were able to obtain the correct final results, while neglecting the convergence generating procedure.

Importantly, the shift \EQref{eqn:convtrafobb} has to be in accordance with the contour defined in \cref{eqn:contourgaussian}. Thus, the choice of the contour fixes the shape of the convergence generating factor. This point is rather crucial, because we are also forced to choose our saddle point in such a way that it is reachable by continuos deformation of the contours in \cref{eqn:contourgaussian,eqn:convtrafobb}. Thus, the Boson-Boson block of the diagonal saddle point solution has to be
\begin{equation}
	\left(\sigma_D^0\right)_{\text{BB}} = \frac{E}{16\pi k} \mathds{1}_4 + \frac{i \Delta}{16\pi k} \diag\left(\mathds{1}_2,-\mathds{1}_2\right) ,
\end{equation} 
where the imaginary parts are on opposite sides of the real axis, \textit{cf.} \cref{eqn:contourgaussian}.
Furthermore, $\widetilde{L}_\text{BB}$, in \cref{eqn:convfac}, guarantees that the exponent is negative but does not influence the relative signs of the terms.

We are now in the position to understand why the transformations exchanging the anticommuting variables of $K$ are not allowed to be non-com\-pact. We show that it is impossible to choose non-compact transformations exchanging the anticommuting variables, have the integrals converge and obtain a valid saddle point solution by continous deformation of the contour. Let us assume that we construct $\Xi^\pm$ in such a way that the anticommuting terms in $K$ have opposite sign. Thus, the transformations exchanging them have to be non-compact. As a direct consequence, the transformations $T$ leave an altered metric $\widetilde{L}\myprime = \diag\left(\mathds{1}_2, -\mathds{1}_2, \mathds{1}_2, -\mathds{1}_2\right) = L$, in Boson-Fermion block notation, invariant and the integral \cref{eqn:hubstratrafo} is no longer convergent again, see the discussion above. Thus, we are forced to introduce an additional factor, similar to \cref{eqn:convfac}, to render our integrals convergent once more. This factor has to be 
\begin{equation}
	\exp\left(\widetilde{\alpha} \frac{N}{4} \str \sigma \widetilde{L}\myprime_{\text{FF}}\right),
\end{equation}
where $\widetilde{L}\myprime_{\text{FF}}=\diag(0_4,\mathds{1}_2,-\mathds{1}_2)$ is the projection of the altered metric $\widetilde{L}\myprime$ onto the Fermion-Fermion block. We note that the lack of an imaginary unit in the exponent is due to the Wick rotation of the Fermion-Fermion block necessary to make $\str\sigma^2$ positive for Hermitian supermatrices. We stress this fact, because it is highly relevant for the discussion at hand.

Proceeding in the same fashion as before, we have to shift the Fermion-Fermion-block into the complex plane by an infinitesimal increment, in the same fashion as in \cref{eqn:convtrafobb}. We recall that the solution to the saddle point equation is given by \cref{eqn:saddlepointeq} and, as a consequence of the Wick rotation for the Fermion-Fermion block, the solutions to saddle point equation are
\begin{equation}
	i x = \frac{E}{16\pi k} \pm \frac{i \Delta}{16\pi k}.
\end{equation}
Consequently, the imaginary parts of the Fermion-Fermion block of $\sigma_D^0$ all lie on the same side of the real axis, as opposed to the bosonic part where they are on different sides. Thus, it is impossible to reconcile a contour of the form \EQref{eqn:convtrafobb}, which is necessary to make the non-compact integrals convergent, and the saddle point contour through a continuous deformation. This is not a consequence of $\widetilde{L}\myprime_{\text{FF}}$, since the different signs of the $[1,1]$ block and $[2,2]$ block contribution in \cref{eqn:fullconvexp} are due to the structure of the group of transformations $T$, and $\widetilde{L}\myprime_{\text{FF}}$ only determines the global sign of the exponent. 

Therefore, it is not permissible to choose $\Xi^\pm$ in such a way that non-compact transformations exchange the anticommuting variables, since there does not exist a contour that both renders the integrals convergent and accounts for all saddle points. This result is rather intuitive, as the integrals over anticommuting variables, which allow for a representation in superspace, are never divergent. Regardless, we appreciate that it is also possible to derive this within our supersymmetric formalism. Thus, we cleared the sign ambiguity, since the transformations $\Xi^\pm$, and, consequently the metric $\widetilde{L}$ is fixed by the above reasoning.

\subsection{Choosing the Correct Saddle Point}
\label{sec:CorrectSaddlePoint}
Unlike for the Boson-Boson block, we do not have a restriction on the Ferm\-ion-Fermion block of the saddle point arising from a convergence argument. Thus, besides the saddle point in \cref{eqn:diagonalsaddlepoint} there also exist the equally permissible solutions in $[1,2]$ block notation
\begin{align}\label{eqn:possibleDiagonalSaddlePoint}
	\sigma_D^0=& \frac{E}{16\pi k} \mathds{1}_8 + \frac{i\Delta}{16\pi k} \diag\left(\mathds{1}_2,-\mathds{1}_2,-\mathds{1}_2,-\mathds{1}_2\right),\notag\\ 
	\sigma_D^0=& \frac{E}{16\pi k} \mathds{1}_8 + \frac{i\Delta}{16\pi k} \diag\left(\mathds{1}_2,\mathds{1}_2,-\mathds{1}_2,\mathds{1}_2\right) ,
\end{align}
due to the compactness of the Fermion-Fermion block. The replacement $L$ by $-L$ is included in \cref{eqn:diagonalsaddlepoint}, since it is reachable by transformation of the diagonal saddle point. Conversely, the other two cases above are not reachable by means of the transformations $T\in\text{UOSp}(2,2\vert 4)$. Thus, the full saddle point approximation is a sum of the integral considered in \cref{subsec:SaddlePointApprox} and equivalent integrals for the other two diagonal solutions in \cref{eqn:possibleDiagonalSaddlePoint}. However, their contributions are of lower order in $N$ and therefore vanish in the limit $N\to\infty$. We show this for the first saddle point in \EQref{eqn:possibleDiagonalSaddlePoint}, the second one follows equivalently.

We consider the diagonal solution
\begin{equation}
	\sigma_D^0= \frac{E}{16\pi k} \mathds{1}_8 + \frac{i\Delta}{16\pi k} \diag\left(\mathds{1}_2,-\mathds{1}_2,-\mathds{1}_2,-\mathds{1}_2\right).
\end{equation}
Here, we solely focus on the massive modes. The contribution of the massive modes is of the form
\begin{equation}
	\exp\left(- \tilde{r} \str \left(\delta P^\prime\right)^2 + 2\tilde{r}^2 \str \left(\sigma_D^0 \delta P^\prime\right)^2\right),
\end{equation}
where $\delta P^\prime = \sqrt{N} R \delta P R^{-1}$. Only the $[1,1]$ block differs from our prior analysis in \cref{subsec:SaddlePointApprox}, as such we neglect the other terms and have
\begin{align}\label{eqn:deltaPnoanticomm}
	&\tilde{r} \str \left(\delta P_{11}^\prime\right)^2 - 2\tilde{r}^2 \str \left(\left[\frac{E}{16\pi k}\mathds{1}_4 + \frac{i\Delta}{16\pi k} \left(\sigma^{(3)}\otimes \mathds{1}_2\right)\right] \delta P_{11}^\prime\right)^2 \\
	=& \left(\frac{(8\pi k)^2}{2v^2}-\frac{(8\pi k)^2}{8v^4} E^2\right) \str\left\{\left(\delta P_{11}^\prime\right)^2 + \left(\left(\sigma^{(3)}\otimes \mathds{1}_2\right) \delta P_{11}^\prime\right)^2 \right\} \notag\\
	&- 2 i (8\pi k)^2 \Delta E \str \left(\sigma^{(3)}\otimes \mathds{1}_2\right) \left(\delta P_{11}^\prime\right)^2 .
\end{align}
Neither of the terms contains any anticommuting variables, which means, unless we get additional contributions from the volume element $\dd[\delta P]$, the integrals over the anticommuting variables vanish and the saddle point does not contribute to the characteristic function. 

We turn to the discussion in Appendix \ref{app:jacobgoldstoneheavy} and see that $\sigma_D^0$ and $\dd\delta P$ no longer commute, since $[\sigma_D^0,R]\neq 0$. Thus, the third term in \cref{eqn:limitElemntaryLengthElement} does not trivially vanish. Instead, it is a function of only the diagonal elements of $P=R\myprime^{-1} P_D R\myprime$ 
\begin{equation}\label{eqn:elementaryLengthConvergence}
	\str \dd \delta P \left[\dd T_0 T_0^{-1}, P\right] = \str \dd \delta P \left[\dd T_0 T_0^{-1}, P_D\right],
\end{equation}
where we absorbed the transformations $R, R\myprime$ into the differentials which leaves the Berezinian invariant. As is evident from the commutator in \linebreak Eq. \EQref{eqn:elementaryLengthConvergence}, the Berezinian only depends on the differences of the elements of $P_D$. The leading order of $P_D$ is given by the saddle point \EQref{eqn:possibleDiagonalSaddlePoint} and as the contributions of the $[2,2]$ block to \cref{eqn:elementaryLengthConvergence} vanish, see Appendix \ref{app:jacobgoldstoneheavy}. However, as the $[1,1]$ block of $\sigma_D^0$ is not proportional the identity, the supertrace \cref{eqn:elementaryLengthConvergence} has non-zero contributions, unlike for the saddle point in \cref{eqn:diagonalsaddlepoint}. Therefore, we have to determine if the first non-zero order contains anticommuting variables. To that end we recall
\begin{equation}
	R P R^{-1} = \sigma_D^0 + \sqrt{N}^{-1} \delta P\myprime .
\end{equation}
As $P$ is block-diagonal, we only focus on the contributions of $P_{11}$. The $[1,1]$ block of $\sigma_D^0$ is, up to a shift in the energy and prefactors, given by $\diag\left(\mathds{1}_2,-\mathds{1}_2\right)$. Thus, we must pay attention to this degeneracy when applying perturbation theory to determine the contributions of $\delta P\myprime$ to the diagonal elements. The first order of a degenerate perturbative expansion does not involve the blocks of the matrix responsible for transitions between different "states". In this case, these blocks are the Boson-Fermion and Fermion-Boson blocks. Thus, the first order corrections are given in terms of the entries of the Boson-Boson and Fermion-Fermion block of $\delta P\myprime_{11}$ only. Anticommuting variables enter in the next order, which scales with $1/N$. Therefore, we conclude that in the leading order in $N$ the anticommuting entries of $P_{11}$ do not appear in the Berezinian and the integral over the anticommuting variables vanishes. This is in accordance with the results for the \emph{GOE} \cite{VWZ1985} and for the tenfold way \cite{Zirnbauer1996}. Hence, in the limit $N\to\infty$ the saddle points \EQref{eqn:possibleDiagonalSaddlePoint} do not contribute to the characteristic function.

\section{Derivation of the Parametrisation of $Q$}
\label{app:parametrisation}
This appendix is structured as follows we will first determine the form of $T_0$ and then diagonalize the different blocks of $T_0$ simultaneously and end with a discussion of the integration domain.

As the transformations $T_0$ are non-trivial transformations of the saddle point, their generators only have off-diagonal blocks in the $[1,2]$ block notation, as opposed to the generators of the transformations $R$. It follows from the form of the generators in \cref{eqn:GeneratorT0} that 
\begin{equation}\label{eqn:BlockFormT0}
	T_0 = \begin{bNiceMatrix}
		\sqrt{\mathds{1}_4 + t_{12} t_{21}} & t_{12} \\
		t_{21} & \sqrt{\mathds{1}_4 + t_{21} t_{12}} 
	\end{bNiceMatrix}, \quad \text{with} \quad t_{21} = \left(\tau^{(3)} \otimes \mathds{1}_2\right) t_{12}^\dagger, 
\end{equation}
when we take into account that the transformations are non-compact unitary, meaning $T_0^\dagger \widetilde{L} T_0 = \widetilde{L}$. Furthermore, the matrix $T_0$ is fully determined by the block $t_{12}$, and it is therefore sufficient to determine and diagonalise $t_{12}$ to find a diagonalisation of the blocks of $T_0$.

The supermatrix $t_{12}$ is determined by the conditions originating from $C T_0^\star C^\trans = T_0$ 
\begin{equation}
	\begin{bNiceMatrix}
		Y & 0 \\
		0 & X
	\end{bNiceMatrix} t_{12}^\star \begin{bNiceMatrix}
		Y & 0 \\
		0 & X
	\end{bNiceMatrix} = t_{12} .
\end{equation}
Solving these equations, we obtain
\begin{equation}\label{eqn:t12}
	t_{12} = \begin{bNiceMatrix}
		i z & i w^\star & \eta_2 & \rho_2 \\
		-i w & i z^\star & \rho_2^\star & \eta_2^\star \\
		\eta_1 & -\rho_1^\star & a & b \\
		\rho_1 & -\eta_1^\star & b^\star & a^\star
	\end{bNiceMatrix} .
\end{equation}
We find that this supermatrix resembles the form of $t_{12}$ investigated in Ref. \cite{VWZ1985}, which becomes even clearer when applying the transformation $(\mathds{1}_2 \oplus \mathfrak{v})$
\begin{equation}\label{eqn:t12tilde}
	\widetilde{t}_{12} = \left(\mathds{1}_2 \oplus \mathfrak{v}\right)^\dagger t_{12} \left(\mathds{1}_2 \oplus \mathfrak{v}\right) = \begin{bNiceMatrix}
		i z & i w^\star & -\widetilde{\eta}_2^\star & -\widetilde{\rho}_2^\star \\
		-i w & i z^\star & \widetilde{\eta}_2 & \widetilde{\rho}_2 \\
		i \widetilde{\eta}_1 & i \widetilde{\eta}_1^\star & \widetilde{a} & \widetilde{b} \\
		i \widetilde{\rho}_1 & i \widetilde{\rho}_1^\star & \widetilde{c} & \widetilde{d}
	\end{bNiceMatrix} .
\end{equation}
The supermatrix $\widetilde{t}_{12}$ is directly related to the supermatrix $t_{12}$ of \linebreak Ref. \cite{VWZ1985}, and they map onto each other if we exchange Boson-Boson with Fermion-Fermion block, and Boson-Fermion with Fermion-Boson block. However, such a mapping does not exist in terms of a transformation contained in the space of supermatrices. Therefore, we diagonalise $\widetilde{t}_{12}$ by a procedure similar to the one of Ref. \cite{VWZ1985}. 

We consider the matrices $\widetilde{\alpha}_1=\widetilde{t}_{12} \widetilde{t}_{21}$ and $\widetilde{\alpha}_2=\widetilde{t}_{21} \widetilde{t}_{12}$, which fulfil
\begin{align}\label{eqn:symmetriesalpha}
	\widetilde{\alpha}_1^\dagger = \widetilde{\alpha}_1, \quad	\widetilde{\alpha}_2^\dagger = \left(\tau^{(3)}\otimes\mathds{1}_2\right) \widetilde{\alpha}_2 \left(\tau^{(3)}\otimes\mathds{1}_2\right), \quad \widetilde{\alpha}_j = \widetilde{C}_j \widetilde{\alpha}_j^\star \widetilde{C}_j.
\end{align}
Here, we introduce $\widetilde{C}_1=Y \oplus \mathds{1}_2$ and $\widetilde{C}_2 = -Y \oplus \mathds{1}_2$, which directly follows from the symmetries of $t_{12}$, \textit{cf.} \cref{eqn:tcondition}.
We find the block structure
\begin{equation}\label{eqn:BlocksMatrixAlpha}
	\widetilde{\alpha}_j = \begin{bNiceMatrix}
		f_j \mathds{1}_2 & i^{j-1}\kappa_j^\dagger \\
		i^{j-1}\kappa_j & b_j
	\end{bNiceMatrix},
\end{equation}
where $f_j = f_j^\star$, $b_j=b_j^\dagger=b_j^\trans$ and $\kappa_j^\star Y^\trans = \kappa_j$. Additionally, we observe that the ordinary part, \textit{i.e.} the part containing only commuting variables and not even orders in anticommuting variables, of $f_j$ is positive, $\ord f_j\geq 0$, and conversely the ordinary part of the eigenvalues of $b_j$ are negative, $\ord b_j\leq 0$. The transformations
\begin{align}\label{eqn:defvtilde}
	\widetilde{v}_j =& \exp\left(i^{j-1} \left(\widetilde{Y} - (-1)^j \frac{\widetilde{Y}^3}{3}\right)\right), \notag\\ 
	\widetilde{Y}_j =& \begin{bNiceMatrix}
		0 & - \widetilde{\xi}^\dagger_j \\
		\widetilde{\xi}_j & 0
	\end{bNiceMatrix}, \quad \widetilde{\xi}_j = \begin{bNiceMatrix}
		\alpha_1 & \alpha_1^\star \\
		\beta_1 & \beta_1^\star
	\end{bNiceMatrix}
\end{align}
block-diagonalise the supermatrix $\widetilde{\alpha}_j$. The anticommuting variables $\alpha_j, \beta_j$ are functions of the entries of $\widetilde{\alpha}_j$, which, if desired, are calculated similarly to Ref. \cite{VWZ1985}. We refrain from giving an explicit dependence, as it is not necessary for our further calculations. We also note that $\widetilde{\xi}_j$ is chosen in such a way that it also fulfils $\widetilde{\xi}_j Y^\trans = \widetilde{\xi}_j$, to embed the correct symmetries onto the transformation $\widetilde{v}_j$. The transformed $\widetilde{\alpha}_j$ take the form
\begin{equation}\label{eqn:BlockDiagonalAlphaMatrix}
	\widetilde{\alpha}_j\myprime = \widetilde{v}_j \widetilde{\alpha}_j \widetilde{v}_j^{-1} = \begin{bNiceMatrix}
		f_j\myprime \mathds{1}_2 & 0 \\
		0 & b_j\myprime
	\end{bNiceMatrix},
\end{equation}
where $f_j\myprime, b_j\myprime$ have the same properties as above, except they also contain even orders in the anticommuting variables contained in the matrices $\kappa_j$, \textit{cf.} \cref{eqn:BlocksMatrixAlpha}. That the transformations in \cref{eqn:defvtilde} perform the claimed block-diagonalisation is best seen by applying the inverse transformation to $\widetilde{\alpha}_j\myprime$ and recovering a supermatrix that fulfils the conditions \cref{eqn:symmetriesalpha}. 

The upper left block of $\widetilde{\alpha}_j\myprime$ in \cref{eqn:BlockDiagonalAlphaMatrix} is already diagonal, and we do not need to apply any further transformations. Since $b_j\myprime$ is real symmetric, we diagonalise it by an $O(2)$ transformation
\begin{equation}
	R(\phi_j) = \begin{bNiceMatrix}
		\cos\phi_j & \sin\phi_j \\
		-\sin\phi_j & \cos\phi_j
	\end{bNiceMatrix},
\end{equation}
and we have
\begin{equation}\label{eqn:diagalpha}
	\widetilde{u}_j \widetilde{\alpha}_j \widetilde{u}_j^{-1} = \diag\left(f_j\myprime,f_j\myprime,b_{j,1}\myprime,b_{j,2}\myprime\right),
\end{equation}
where $\widetilde{u}_j = \widetilde{O}_j \widetilde{v}_j$, $\widetilde{O}_j = \diag\left(\mathds{1}_2, R(\phi_j)\right)$ and $b_{j,n}\myprime, \, n=1,2$ are the eigenvalues of $b_j\myprime$. As both $\widetilde{\alpha}_1$ and $\widetilde{\alpha}_2$ are square supermatrices, their characteristic polynomials coincide, and they must have the same zeroes and therefore singular values. We choose the angles $\phi_j$ in such a way that $b_{1,n}\myprime=b_{2,n}\myprime$, which means the right-hand side of \cref{eqn:diagalpha} is the same for $j=1$ and $j=2$.

We turn back to the supermatrices $\widetilde{t}_{12}$ and $\widetilde{t}_{21}$, and define
\begin{align}
	\widehat{t}_{12} =& \widetilde{u}_1 \widetilde{t}_{12} \widetilde{u}_2^{-1} \notag\\
	\widehat{t}_{21} =& \widetilde{u}_2 \widetilde{t}_{21} \widetilde{u}_1^{-1},
\end{align}
which commute, as their products are given by the right-hand side of \linebreak\cref{eqn:diagalpha}. Thus, we diagonalise them simultaneously. The supermatrices $\widehat{t}_{12}$ and $\widehat{t}_{21}$ only differ from a diagonal matrix in the Boson-Boson block due to the degeneracy, since \cref{eqn:diagalpha} is diagonal. Comparing the explicit form of $\widetilde{t}_{12}$ and the Boson-Boson block of $\widehat{t}_{12}$, we find that up to some nilpotent terms it has the form
\begin{equation}
	\begin{bNiceMatrix}
		i z & i w^\star \\
		-i w & i z^\star
	\end{bNiceMatrix} = i \mu U,
\end{equation}
with $\mu=\sqrt{z z^\star+w w^\star}\geq 0$ and a special unitary transformation $U\in \text{SU}(2)$. Collecting all of the above considerations, we arrive at 
\begin{align}\label{eqn:diagonaltildet12}
	\widetilde{t}_{12} =& \widetilde{u}_1^{-1} \begin{bNiceMatrix}
		i \mu U & 0 & 0 \\
		0 & \mu_1 & 0  \\
		0 & 0 & \mu_2
	\end{bNiceMatrix} \widetilde{u}_2, \notag\\
	\widetilde{t}_{21} =& - \widetilde{u}_2^{-1} \begin{bNiceMatrix}
		i \mu U^\dagger & 0 & 0 \\
		0 & \mu_1 & 0  \\
		0 & 0 & \mu_2
	\end{bNiceMatrix} \widetilde{u}_1 .
\end{align}
We directly infer the form of $t_{12}$ from \cref{eqn:t12tilde}, which results in
\begin{equation}\label{eqn:DiagonalizationT12}
	t_{12} = \left(\mathds{1}_2 \oplus \mathfrak{v}\right) \widetilde{t}_{12} \left(\mathds{1}_2 \oplus \mathfrak{v}\right)^\dagger = u_1^{-1} \widehat{U} \begin{bNiceMatrix}
		i \mu \mathds{1}_2 & 0 & 0 \\
		0 & \frac{\mu_1+\mu_2}{2} & \frac{\mu_1-\mu_2}{2} \\
		0 & \frac{\mu_1-\mu_2}{2} & \frac{\mu_1+\mu_2}{2} 
	\end{bNiceMatrix} u_2
\end{equation}
and accordingly for $\widetilde{t}_{21}$. Here, the transformations $u_j$ are
\begin{align}\label{eqn:Defu1u2}
	u_j =& \left(\mathds{1}_2 \oplus \mathfrak{v}\right) \widetilde{u}_j \left(\mathds{1}_2 \oplus \mathfrak{v}\right)^\dagger = O_j v_j, \notag\\
	O_j =& \diag\left(\mathds{1}_2, e^{-i\phi_j}, e^{i\phi_j}\right) .
\end{align}
The transformations $v_j$ are generated by the matrices
\begin{equation}\label{eqn:xidef}
	\xi_j = \begin{bNiceMatrix}
		\mu_j & \nu_j^\star \\
		\nu_j & \mu_j^\star
	\end{bNiceMatrix} 
\end{equation}
fulfilling
\begin{equation}\label{eqn:symmetryxi}
	X \xi_j Y = \xi_j^\star
\end{equation}
in accordance with the symmetries of $t_{12}$. In the following, we decide to use Efetov's angles \cite{Efetov1997} 
\begin{align}\label{eqn:DefEfetovAngles}
	\mu = \sinh\frac{\theta}{2}, \quad \mu_1 = \sin\frac{\theta_1 + \theta_2}{2}, \quad \mu_2 = \sin\frac{\theta_1 - \theta_2}{2} ,
\end{align}
instead of the variables used in Ref. \cite{VWZ1985}, as they are more convenient to work with. 

Regarding the domain of the parametrisation, we know that $\theta>0$ and $\widehat{U}\in\text{SU}(2)$ from the derivations above, but have yet to specify the domain of the variables $\theta_1$ and $\theta_2$. For this, we simply compare the derived form of $t_{12}$ with its diagonalised form. The Fermion-Fermion blocks are
\begin{align}
	&\begin{bNiceMatrix}
		a & b \\
		b^\star & a^\star
	\end{bNiceMatrix}, \notag\\
	&\begin{bNiceMatrix}
		\cos\dfrac{\theta_2}{2} \sin\dfrac{\theta_1}{2} e^{-i(\phi_2-\phi_1)} & \cos\dfrac{\theta_1}{2} \sin\dfrac{\theta_2}{2} e^{i(\phi_2+\phi_1)}\\
		\cos\dfrac{\theta_1}{2} \sin\dfrac{\theta_2}{2} e^{-i(\phi_2+\phi_1)} & \cos\dfrac{\theta_2}{2} \sin\dfrac{\theta_1}{2} e^{i(\phi_2-\phi_1)}
	\end{bNiceMatrix} ,
\end{align}
only considering the ordinary part of the singular value decomposition.
Using the polar decomposition for the complex numbers $a=r_a e^{i\phi_a}$ and $b=r_b e^{i\phi_b}$, we see that both $\cos\left(\theta_2/2\right) \sin\left(\theta_1/2\right)$ and $\sin\left(\theta_2/2\right) \cos\left(\theta_1/2\right)$ have to be positive, as the phase contribution is already accounted for by the orthogonal degrees of freedom. This is only possible if we restrict the angles to $\theta_1, \theta_2 \in[0,\pi]$.

Interestingly, there is still further double counting with the above restriction. We turn to the singular values $\sin\left(\theta_1+\theta_2\right)/2$ and $\sin\left(\theta_1-\theta_2\right)/2$, and divide the interval $[0,\pi]^2$ into $[0,\pi/2]^2$, $[0,\pi/2]\times [\pi/2,\pi]$, $[\pi/2,\pi]\times [0,\pi/2]$ and $[\pi/2,\pi]^2$. Starting with the last interval, we employ the change of variables $\theta_j=\pi- \widetilde{\theta}_j$, where $\widetilde{\theta}_j\in [0,\pi/2]$, and we observe
\begin{equation}
	\sin\frac{\theta_1+\theta_2}{2} = \sin\frac{\widetilde{\theta}_1+\widetilde{\theta}_2}{2} \quad \text{and} \quad \sin\frac{\theta_1-\theta_2}{2} = - \sin\frac{\widetilde{\theta}_1-\widetilde{\theta}_2}{2} .
\end{equation}
To show this explicitly, one can change to centre and difference coordinates, while minding their domain. Clearly, the first singular value is the same as for the interval $[0,\pi/2]^2$. Although not as obvious, the second singular value is the same as well. This is, because $\theta_1-\theta_2$ is defined on an interval with inversion symmetry around zero, such that the image is the same as for $[0,\pi/2]^2$. We perform similar considerations for the remaining two intervals and find, for $[\pi/2,\pi]\times [0,\pi/2]$,
\begin{equation}
	\sin\frac{\theta_1+\theta_2}{2} = \cos\frac{\widetilde{\theta}_1-\theta_2}{2} \quad \text{and} \quad \sin\frac{\theta_1-\theta_2}{2} = \cos\frac{\widetilde{\theta}_1+\theta_2}{2},
\end{equation}
and similarly, for $[0,\pi/2]\times [\pi/2,\pi]$,
\begin{equation}
	\sin\frac{\theta_1+\theta_2}{2} = \cos\frac{\theta_1-\widetilde{\theta}_2}{2} \quad \text{and} \quad \sin\frac{\theta_1-\theta_2}{2} = - \cos\frac{\theta_1+\widetilde{\theta}_2}{2} .
\end{equation}
Again, the first singular values coincide in both cases, while the second ones are the same up to a sign. Keeping in mind that the cosine has an inversion symmetry around $\pi/2$ and $\theta_1+\widetilde{\theta}_2$ is defined on such an interval, the image is the same as for $[\pi/2,\pi]\times [0,\pi/2]$.

We conclude that, to avoid double counting, the angles $\theta_1, \theta_2$ have to be restricted to the interval $[0,\pi]\times [0,\pi/2]$. The choice of which angle is defined on which interval is arbitrary, since the two angles are interchangeable. In the following, we let both angles be in $[0,\pi]^2$ and compensate the double counting by an additional factor $1/2$. This might seem arbitrary, but it is necessary to carry out the integrals over the saddle point manifold, due to some technicalities involving integration in superspace.

Such considerations are not necessary for the \emph{GOE}, because in that case, the degenerate singular value $\theta$ is compact, whereas $\theta_1$ and $\theta_2$ are non-compact. The restrictions derived above depend on the fact that $\theta_1$ and $\theta_2$ have a compact domain where, we can apply trigonometric identities. This is not the case if the variables are non-compact. Additionally, since $\theta$ is degenerate, simply restricting $\theta$ to $[0,\pi]$ is sufficient for the \emph{GOE}.

We return to the matrix $Q$, as it was our original goal to find an explicit parametrisation. We insert \cref{eqn:DiagonalizationT12} into the block form of $T_0$, see \cref{eqn:BlockFormT0}, and get
\begin{equation}
	T_0 = \mathcal{U}^{-1} \begin{bNiceMatrix}
		\sqrt{\mathds{1}_4 - M^2} & M \\
		- M & \sqrt{\mathds{1}_4 - M^2}
	\end{bNiceMatrix} \mathcal{U} ,
\end{equation}
where we introduce
\begin{align}\label{eqn:DefinitionU}
	\mathcal{U} =& \begin{bNiceMatrix}
		\widehat{U}^\dagger u_1 & 0 \\
		0 & u_2
	\end{bNiceMatrix} , \notag\\
	M =& \begin{bNiceMatrix}
		i \sinh\frac{\theta}{2} \mathds{1}_2 & 0 & 0 \\
		0 & \cos\frac{\theta_2}{2} \sin\frac{\theta_1}{2} & \cos\frac{\theta_1}{2} \sin\frac{\theta_2}{2} \\
		0 & \cos\frac{\theta_1}{2} \sin\frac{\theta_2}{2} & \cos\frac{\theta_2}{2} \sin\frac{\theta_1}{2}
	\end{bNiceMatrix} = \sin \frac{\widehat{\theta}}{2}, \notag\\
	\widehat{\theta} =& \begin{bNiceMatrix}
		i \theta \mathds{1}_2 & 0 & 0 \\
		0 & \theta_1 & \theta_2 \\
		0 & \theta_2 & \theta_1
	\end{bNiceMatrix},
\end{align}
with the transformations $u_j$ and the Efetov angles defined in \cref{eqn:Defu1u2} and \cref{eqn:DefEfetovAngles}, respectively. As a consequence of the restriction of $(\theta_1, \theta_2)$ to the interval $[0,\pi]\times[0,\pi/2]$, it follows that
\begin{equation}
	\sqrt{\mathds{1}_4 - M^2} = \cos\frac{\widehat{\theta}}{2},
\end{equation}
where no absolute values appear, as the matrix is always positive semi-definite for $\theta_1+\theta_2\leq \pi/2$ and negative definite for $\theta_1 + \theta_2 > \pi/2$, in which case the additional negative sign is pointwise absorbed into the transformations $O_j$, which leaves the integration invariant due to the $2\pi$-periodicity. Thus, we have successfully block-diagonalised the transformations $T_0$
\begin{equation}
	T_0 = \mathcal{U}^{-1} \begin{bNiceMatrix}
		\cos\frac{\widehat{\theta}}{2} & \sin\frac{\widehat{\theta}}{2} \\
		\sin\frac{\widehat{\theta}}{2} & -\cos\frac{\widehat{\theta}}{2}
	\end{bNiceMatrix} \mathcal{U} .
\end{equation}
Inserting this into the definition of the matrix $Q=-i T_0 L T_0$ gives
\begin{equation}\label{eqn:ParametrizationQ}
	Q = \mathcal{U}^{-1} \begin{bNiceMatrix}
		\cos\widehat{\theta} & \sin\widehat{\theta} \\
		\sin\widehat{\theta} & -\cos\widehat{\theta}
	\end{bNiceMatrix} \mathcal{U} .
\end{equation}
It is important to keep in mind that we neglect the fact that, as a consequence of the diagonalisation procedure, all commuting variables also contain even orders of the anticommuting variables in \cref{eqn:t12}. Neglecting these and only considering the ordinary part of the commuting values is, in general, not correct and results in the Efetov-Wegner Terms discussed in \cref{app:efetovwegner}.  

\section{Determining the volume element $\dd\mu(Q)$}
\label{app:volumelemegoldstone}
As seen in \ref{app:jacobgoldstoneheavy} the volume element factorizes into a contribution from the heavy modes and the Goldstone modes. As the non-linear sigma model contains an integration over the Goldstone modes we want to determine the corresponding volume element for the parametrization derived in \ref{app:parametrisation}. We do this again by means of the invariant length element $\str\dd Q^2$ where $Q=-i \mathcal{U}^{-1} Q_0 \mathcal{U}$.
Inserting the definition we find for the invariant length element
\begin{align}
	\str \dd Q^2 =& - \str \Biggl(\dd Q_0^2 - 2 \delta\mathcal{U}_1^2 - 2 \delta\mathcal{U}_2^2 + 2 \left(\delta\mathcal{U}_1 \cos\widehat{\theta}\right)^2 + 2 \left(\delta\mathcal{U}_2 \cos\widehat{\theta}\right)^2 \notag\\
	+& 4 \delta\mathcal{U}_1 \sin\widehat{\theta} \delta\mathcal{U}_2 \sin\widehat{\theta}\Biggr)
\end{align}
where $\delta\mathcal{U}_j = \mathcal{U}\dd\mathcal{U}^{-1} = - \dd\mathcal{U} \mathcal{U}^{-1}$. These contributions can be broken down further into those containing commuting and anticommuting blocks of $\delta\mathcal{U}_j$ similar to \cite{Efetov1997}
\begin{equation}
	\str \dd Q^2 = - \str\dd Q_0^2 + \left[\str\dd Q^2\right]_{\text{commuting}} + \left[\str\dd Q^2\right]_{\text{anticommuting}}
\end{equation}
with
\begin{align}
	\left[\str\dd Q^2\right]_{\text{commuting}} =& \tr \Biggl(
	2 \left(\delta\mathcal{U}_{1}^\text{BB}\right)^2 + 2 \left(\delta\mathcal{U}_{2}^\text{BB}\right)^2 - 2 \left(\delta\mathcal{U}_{1}^\text{FF}\right)^2 - 2 \left(\delta\mathcal{U}_{2}^\text{FF}\right)^2 \notag\\
	-& 2 \left(\delta\mathcal{U}_{1}^\text{BB} \cos\widehat{\theta}_{\text{BB}}\right)^2 - 2 \left(\delta\mathcal{U}_{2}^\text{BB} \cos\widehat{\theta}_{\text{BB}}\right)^2 \notag\\
	+& 2 \left(\delta\mathcal{U}_{1}^\text{FF} \cos\widehat{\theta}_{\text{FF}}\right)^2 + 2 \left(\delta\mathcal{U}_{2}^\text{FF} \cos\widehat{\theta}_{\text{FF}}\right)^2 \notag\\
	-& 4 \delta\mathcal{U}_{1}^\text{BB} \sin\widehat{\theta}_{\text{BB}} \delta\mathcal{U}_{2}^\text{BB} \sin\widehat{\theta}_{\text{BB}} + 4 \delta\mathcal{U}_{1}^\text{FF} \sin\widehat{\theta}_{\text{FF}} \delta\mathcal{U}_{2}^\text{FF} \sin\widehat{\theta}_{\text{FF}}
	\Biggr)
\end{align}
and
\begin{align}
	\left[\str\dd Q^2\right]_{\text{anticommuting}} =& \tr \Biggl(
	4 \delta\mathcal{U}_{1}^\text{BF} \delta\mathcal{U}_{1}^\text{FB} + 4 \delta\mathcal{U}_{2}^\text{BF} \delta\mathcal{U}_{2}^\text{FB} \notag\\
	-& 4 \delta\mathcal{U}_{1}^\text{BF} \cos\widehat{\theta}_{\text{FF}} \delta\mathcal{U}_{1}^\text{FB} \cos\widehat{\theta}_{\text{BB}} - 4 \delta\mathcal{U}_{2}^\text{BF} \cos\widehat{\theta}_{\text{FF}} \delta\mathcal{U}_{2}^\text{FB} \cos\widehat{\theta}_{\text{BB}} \notag\\
	-& 8 \delta\mathcal{U}_{1}^\text{BF} \sin\widehat{\theta}_{\text{FF}} \delta\mathcal{U}_{2}^\text{FB} \sin\widehat{\theta}_{\text{BB}}
	\Biggr) .
\end{align}
For the singular values we find 
\begin{equation}
	- \str \dd Q_0^2 = 4 \left(\dd\theta^2 - \dd\theta_1^2 - \dd\theta_2^2\right)
\end{equation}
and for the remaining two terms we explicitly calculate the blocks of $\delta\mathcal{U}_j$. These calculations are greatly simplified by performing certain substitutions, as these are not straightforward we list them. For the anticommuting blocks of $\delta\mathcal{U}_j$ we get
\begin{equation}
	\delta\mathcal{U}_1^{\text{BF}} = - \dd\widehat{\xi}_1^\dagger, \quad \delta\mathcal{U}_1^{\text{FB}} = \dd\widehat{\xi}_1, \quad 	\delta\mathcal{U}_2^{\text{BF}} = -i \dd\widehat{\xi}_2^\dagger, \quad 	\delta\mathcal{U}_2^{\text{FB}} = i \dd\widehat{\xi}_2
\end{equation}
where 
\begin{align}
	\dd\widehat{\xi}_1 =& \diag\left(e^{-i \phi_1},e^{i \phi_1}\right) \Theta_1 \dd\xi_1 U \quad \text{with} \quad \Theta_1 = \begin{bNiceMatrix}
		-1 + \dfrac{1}{2} \mu_1\mu_1^\star \nu_1\nu_1^\star & \mu_1 \nu_1^\star \\
		-\mu_1^\star \nu_1 & -1 + \dfrac{1}{2} \mu_1\mu_1^\star \nu_1\nu_1^\star
	\end{bNiceMatrix}\notag\\
	\dd\widehat{\xi}_1 =& \diag\left(e^{-i \phi_2},e^{i \phi_2}\right) \Theta_2 \dd\xi_2 \quad \text{with} \quad \Theta_2 = \begin{bNiceMatrix}
		-1 + \dfrac{1}{2} \mu_2\mu_2^\star \nu_2\nu_2^\star & -\mu_2 \nu_2^\star \\
		\mu_2^\star \nu_2 & -1 + \dfrac{1}{2} \mu_2\mu_2^\star \nu_2\nu_2^\star
	\end{bNiceMatrix} .
\end{align}
The matrices $\Theta_j$ are Hermitian, have unit determinant and fulfil $\Theta_1^\star = X \Theta_1 X$ such that the properties of $\dd\xi_j$ remain unchanged. The additional imaginary units for $j=2$ reflect the non-compact nature of the transformations $v_2$, as presented in the prior appendix. The new variables greatly simplify the calculation of the invariant length element. Writing the independent anticommuting variables as a vector $\dd\vec{\Upsilon}= (\dd\mu_1,\dd\nu_1,\dd\nu_1^\star,\dd\mu_1^\star,\dd\mu_2,\dd\nu_2,\dd\nu_2^\star,\dd\mu_2^\star)$ which we can relate to the hatted variables
\begin{equation}
	\dd\widehat{\vec{\Upsilon}} = \bm{\Theta}\dd\vec{\Upsilon} \quad \text{with} \quad \bm{\Theta} = \left(U^\trans \otimes \left(\diag\left(e^{-i\phi_1},e^{i\phi_1}\right)\Theta_1\right)\right)\oplus \left(\mathds{1}_2 \otimes \left(\diag\left(e^{-i\phi_2},e^{i\phi_2}\right)\Theta_2\right)\right) .
\end{equation}
We then find 
\begin{equation}
	\left[\str\dd Q^2\right]_{\text{anticommuting}} = 4 \dd\vec{\Upsilon}^\dagger \bm{\Theta}^\dagger \begin{bNiceMatrix}
		a & 0 & b & 0 \\
		0 & a & 0 & b \\
		b & 0 & -a & 0 \\
		0 & b & 0 & -a
	\end{bNiceMatrix} \Theta \dd\vec{\Upsilon}
\end{equation}
with the matrices
\begin{align}
	a =& \begin{bNiceMatrix}
		-1+\cos\theta_1 \cos\theta_2 \cosh\theta & -\sin\theta_1 \sin\theta_2 \cosh\theta\\
		-\sin\theta_1 \sin\theta_2 \cosh\theta & -1+\cos\theta_1 \cos\theta_2 \cosh\theta
	\end{bNiceMatrix} \notag\\
	\text{and} \quad b =& \begin{bNiceMatrix}
	- \sin\theta_1 \cos\theta_2 \sinh\theta & - \cos\theta_1 \sin\theta_2 \sinh\theta \\
	- \cos\theta_1 \sin\theta_2 \sinh\theta & - \sin\theta_1 \cos\theta_2 \sinh\theta
	\end{bNiceMatrix} .
\end{align}
For the commuting term we distinguish between contributions from the Boson-Boson- and the Fermion-Fermion block
\begin{equation}
	\left[\str \dd Q^2\right]_{\text{commuting}} = \left[\str \dd Q^2\right]_{\text{commuting}}^{\text{BB}} + \left[\str \dd Q^2\right]_{\text{commuting}}^{\text{FF}}.
\end{equation}
For the Boson-Boson-block we find
\begin{align}
	\left[\str \dd Q^2\right]_{\text{commuting}}^{\text{BB}} =& -2 \sinh^2\theta \tr\left(\delta\mathcal{U}_1^{\text{BB}}-\delta\mathcal{U}_2^{\text{BB}}\right)^2 \notag\\
	=& 8 \sinh^2\theta \tr \left(\left(\mathds{1}_2-i M\right)^{-1} \dd \widetilde{M} \left(\mathds{1}_2 + i M\right)^{-1}\right)^2 \notag\\
	=& 2^4 \sinh^2\theta \frac{\dd \widetilde{m}^2 + \dd\widetilde{m}_1 \dd\widetilde{m}_1^\star}{\left(1+m^2+m_1m_1^\star\right)^2}
\end{align}
where we have introduced the new variables
\begin{align}
	i\dd\widetilde{M} =& i\dd M - \frac{1}{4} \left(\mathds{1}_2 + i M\right) \left(\dd\widetilde{\xi}_1^\dagger \Theta_1 \widetilde{\xi}_1 - \widetilde{\xi}_1^\dagger \Theta_1 \dd\widetilde{\xi}_1\right) \left(\mathds{1}_2 - i M\right) \notag\\
	+& \frac{1}{4} \left(\mathds{1}_2 - i M\right) \left(\widetilde{\xi}_2^\dagger \Theta_2 \dd\widetilde{\xi}_2 - \dd\widetilde{\xi}_2^\dagger \Theta_2 \widetilde{\xi}_2\right) \left(\mathds{1}_2 + i M\right)  
\end{align}
which will significantly aid in the calculation of the total Berizinian as in the new variables there are no terms that couple commuting and anticommuting degrees of freedom. This comes at a cost which we will return to at the end of this appendix.

Similarly, we find for the Fermion-Fermion block 
\begin{align}
	\left[\str\dd Q^2\right]_{\text{commuting}}^{\text{FF}}	=& 2 \tr \Biggl(-\left(\delta\mathcal{U}_1^{\text{FF}}\right)^2-\left(\delta\mathcal{U}_2^{\text{FF}}\right)^2 + \left(\delta\mathcal{U}_1^{\text{FF}} \cos\widehat{\theta}_{\text{FF}}\right)^2 + \left(\delta\mathcal{U}_2^{\text{FF}} \cos\widehat{\theta}_{\text{FF}}\right)^2 \notag\\
	+& 2 \delta\mathcal{U}_1^{\text{FF}} \sin\widehat{\theta}_{\text{FF}} \delta\mathcal{U}_2^{\text{FF}} \sin\widehat{\theta}_{\text{FF}}\Biggr) \notag\\
	=& \begin{bNiceMatrix}
		\dd\widetilde{\phi}_1 \\
		\dd\widetilde{\phi}_2
	\end{bNiceMatrix}^\trans 4 \begin{bNiceMatrix}
		\sin^2\theta_1 + \sin^2\theta_2 & \sin^2\theta_1 - \sin^2\theta_2 \\
		\sin^2\theta_1 - \sin^2\theta_2 & \sin^2\theta_1 + \sin^2\theta_2
	\end{bNiceMatrix} \begin{bNiceMatrix}
		\dd\widetilde{\phi}_1 \\
		\dd\widetilde{\phi}_2
	\end{bNiceMatrix} 
\end{align}
where we have again introduced new angles
\begin{align}
	\dd\widetilde{\phi}_1 =& \dd\phi_1 + \frac{i}{2} \left(\dd\mu_1 \mu_1^\star - \mu_1 \dd\mu_1^\star - \dd\nu_1 \nu_1^\star + \nu_1 \dd\nu_1^\star\right) \notag\\
	\dd\widetilde{\phi}_2 =& \dd\phi_2 - \frac{i}{2} \left(\dd\mu_2 \mu_2^\star - \mu_2 \dd\mu_2^\star - \dd\nu_2 \nu_2^\star + \nu_2 \dd\nu_2^\star\right) .
\end{align}
Collecting all of our results and paying mind to the fact the anticommuting contribution to the invariant length element will contribute inversely to the Berizinian we arrive at 
the Berizinian given in \cref{eqn:volumegoldstone}. However, we have yet justified why it is possible to only use the ordinary part $\widetilde{m}, \widetilde{m}_1$ and $\widetilde{\phi}_j$. We will only discuss this briefly as we carry out a more detailed discussion for the same problem concerning the singular values in \ref{app:efetovwegner} and all the steps outlined can be undertaken here as well. However, in this case the result is much easier, as one would have to express all the variables without tilde by those with tilde there would exist additional anticommuting parts. These parts will yield, as discussed in \ref{app:efetovwegner}, boundary contributions of the integral. However, as the integrand only depends on the diagonal elements of $U$, which are constant on the boundary. The terms cancel, and it is permissible to use the ordinary parts $m, m_1$ and $\phi_j$ instead.

\section{Calculation of the Phase $F_s^{(4)}$}
\label{app:phasecontribution}
Our non-linear sigmal model, derived in \cref{eqn:nonlinearsigma}, includes a phase factor which depends on the elements of the matrix $\rho^{(c)}$. To be able to expand the integral in anticommuting variables and carry out some of the remaining integrals it is necessary to determine these elements explicitly. We do this by means of the saddle point equation and the singular value decomposition and find that 
\begin{equation}
	\rho^{(c)} = - \frac{4\pi k}{v^2} \mathcal{U}^{-1} \left(\mathds{1}_8 - \frac{i\gamma_c}{2v^2} \left(E \mathds{1}_8 + i \Delta Q_0\right)L\right)^{-1} \left(E \mathds{1}_8 + i\Delta Q_0\right) \mathcal{U} .
\end{equation}
We can calculate the inverse and get
\begin{equation}
	\left(\mathds{1}_8 - \frac{i\gamma_c}{2v^2} \left(E \mathds{1}_8 + i \Delta Q_0\right)L\right)^{-1} = d \left(\mathds{1}_8 + \frac{i\gamma_c}{2v^2} L \left(E\mathds{1}_8 - i \Delta Q_0\right)\right)
\end{equation}
with the matrix
\begin{equation}
	d = \left(\left(1+\frac{\gamma_c^2}{v^2}\right) \mathds{1}_8 + \frac{\Delta \gamma_c}{2v^2} \left(L Q_0 + Q_0 L\right)\right)^{-1}.
\end{equation}
At this point it is useful to fully diagonalize the blocks of $Q_0$ which is achieved by means of the transformation $\bm{V}=\diag\left(\mathds{1}_2,\mathfrak{v},\mathds{1}_2,\mathfrak{v}\right)$ with which we have
\begin{equation}
	Q_0\myprime = \bm{V}^\dagger Q_0 \bm{V} = \begin{bNiceMatrix}
		\cos\widetilde{\theta} & \sin\widetilde{\theta} \\
		\sin\widetilde{\theta} & -\cos\widetilde{\theta}
	\end{bNiceMatrix} \quad \text{where} \quad \widetilde{\theta} = \diag\left(i \theta\mathds{1}_2, \theta_1+\theta_2, \theta_1-\theta_2\right)
\end{equation}
and
\begin{equation}
	d\myprime^{-1} = \bm{V}^\dagger d^{-1} \bm{V} =  \mathds{1}_2 \otimes \frac{\Delta\gamma_c}{v^2} \left(g_c^+ \mathds{1}_4 + \diag\left(\cosh\theta \mathds{1}_2,\cos\left(\theta_1+\theta_2\right),\cos\left(\theta_1-\theta_2\right)\right)\right)
\end{equation}
where we have introduced $g_c^+=\left(\gamma_c^2+v^2\right)/\left(\gamma_c\Delta\right)$. With that we can write $\rho^{(c)}$ as
\begin{equation}
	\rho^{(c)} = - \frac{4\pi k}{v^2} \mathcal{U}^{-1} \bm{V} d\myprime  \left(\mathds{1}_8 + \frac{i\gamma_c}{2v^2} L \left(E\mathds{1}_8 - i \Delta Q_0\myprime\right)\right) \left(E \mathds{1}_8 + i\Delta Q_0\myprime\right) \bm{V}^\dagger \mathcal{U} .
\end{equation}
As $F_s^{(4)}$ does not depend on all of $\rho^{(c)}$ but rather only distinct elements which allows us to restrict to certain blocks. These blocks are the bosonic blocks of $\rho^{(c)}_{12}$ and $\rho^{(c)}_{21}$ which are given by
\begin{align}
	\left[\rho^{(c)}_{12}\right]_{\text{BB}} =& -i\frac{4\pi k}{\gamma_c} \Biggl(i\frac{\sinh\theta}{g_c^+ + \cosh\theta} \mathcal{U}_{1,\text{BB}}^{-1} \mathcal{U}_{2,\text{BB}} + \mathcal{U}_{1,\text{BF}}^{-1} \mathfrak{v} \text{ds}\left(\theta_1,\theta_2\right)\mathfrak{v}^\dagger \mathcal{U}_{2,\text{FB}}\Biggr) \notag\\
	\left[\rho^{(c)}_{21}\right]_{\text{BB}} =& -i\frac{4\pi k}{\gamma_c} \Biggl(i\frac{\sinh\theta}{g_c^+ + \cosh\theta} \mathcal{U}_{2,\text{BB}}^{-1} \mathcal{U}_{1,\text{BB}} + \mathcal{U}_{2,\text{BF}}^{-1} \mathfrak{v} \text{ds}\left(\theta_1,\theta_2\right)\mathfrak{v}^\dagger \mathcal{U}_{1,\text{FB}}\Biggr)
\end{align}
where
\begin{equation}
	\text{ds}\left(\theta_1,\theta_2\right)=\diag\left(\frac{\sin\left(\theta_1+\theta_2\right)}{g_c^+ + \cos\left(\theta_1+\theta_2\right)},\frac{\sin\left(\theta_1-\theta_2\right)}{g_c^+ + \cos\left(\theta_1-\theta_2\right)}\right).
\end{equation}
We can now insert the derived parametrization in \ref{app:parametrisation} and determine the phase, given in \cref{eqn:fs4bf} and arrive at
\begin{align}
	F_s^{(4)} = i^s 8k \Biggl(&i \frac{\sinh\theta}{g_b^+ + \cosh\theta}U_{11} \left(1+ \frac{1}{2}\left(\mu_1\mu_1^\star + \nu_1\nu_1^\star\right)+\frac{3}{4}\mu_1\mu_1^\star\nu_1\nu_1^\star\right) \notag\\
	&\times\left(1- \frac{1}{2}\left(\mu_2\mu_2^\star + \nu_2\nu_2^\star\right)+\frac{3}{4}\mu_2\mu_2^\star\nu_2\nu_2^\star\right) \notag\\
	&+ \kappa_1^{(b)} \nu_1^\star\left(1+\frac{\mu_1\mu_1^\star}{2}\right) i\nu_2\left(1-\frac{\mu_2\mu_2^\star}{2}\right) \notag\\
	+& {\kappa_1^{(b)}}^\star \mu_1^\star\left(1+\frac{\nu_1\nu_1^\star}{2}\right) i\mu_2\left(1-\frac{\nu_2\nu_2^\star}{2}\right) \notag\\
	&+ \kappa_2^{(b)} \mu_1^\star\left(1+\frac{\nu_1\nu_1^\star}{2}\right) i\nu_2\left(1-\frac{\mu_2\mu_2^\star}{2}\right) \notag\\
	&+ {\kappa_2^{(b)}}^\star \nu_1^\star\left(1+\frac{\mu_1\mu_1^\star}{2}\right) i\mu_2\left(1-\frac{\nu_2\nu_2^\star}{2}\right)\Biggr) \notag\\
	-(-i)^{s} 8k \Biggl(& i \frac{\sinh\theta}{g_a^+ + \cosh\theta}U_{22} \left(1+ \frac{1}{2}\left(\mu_1\mu_1^\star + \nu_1\nu_1^\star\right)+\frac{3}{4}\mu_1\mu_1^\star\nu_1\nu_1^\star\right) \notag\\
	&-\kappa_1^{(a)} \mu_1\left(1+\frac{\nu_1\nu_1^\star}{2}\right) i\mu_2^\star\left(1-\frac{\nu_2\nu_2^\star}{2}\right) \notag\\
	&- {\kappa_1^{(a)}}^\star \nu_1\left(1+\frac{\mu_1\mu_1^\star}{2}\right) i\nu_2^\star\left(1-\frac{\mu_2\mu_2^\star}{2}\right) \notag\\
	&- \kappa_2^{(a)} \nu_1\left(1+\frac{\mu_1\mu_1^\star}{2}\right) i\mu_2^\star\left(1-\frac{\nu_2\nu_2^\star}{2}\right) \notag\\
	&- {\kappa_2^{(a)}}^\star \mu_1\left(1+\frac{\nu_1\nu_1^\star}{2}\right) i\nu_2^\star\left(1-\frac{\mu_2\mu_2^\star}{2}\right)\Biggr) .
	\label{eqn:fs4saddlepoint}
\end{align}
where 
\begin{align}
	\kappa_1^{(c)} =& \frac{1}{2} e^{i\left(\phi_2-\phi_1\right)} \left(\frac{\sin\left(\theta_1+\theta_2\right)}{g_c^+ + \cos\left(\theta_1+\theta_2\right)} + \frac{\sin\left(\theta_1-\theta_2\right)}{g_c^+ + \cos\left(\theta_1-\theta_2\right)}\right) \notag\\
	\text{and} \quad \kappa_2^{(c)} =& \frac{1}{2} e^{i\left(\phi_2+\phi_1\right)} \left(\frac{\sin\left(\theta_1+\theta_2\right)}{g_c^+ + \cos\left(\theta_1+\theta_2\right)} - \frac{\sin\left(\theta_1-\theta_2\right)}{g_c^+ + \cos\left(\theta_1-\theta_2\right)}\right) .
\end{align}
From \cref{eqn:fs4saddlepoint} we can infer that the commuting part of the phase $F_s^{(4)}$ is given by
\begin{equation}
	\kappa_0 := \left[F_s^{(4)}\right]_{\text{commuting}} = 8k\sinh\theta \left(\frac{i^{s+1}}{g_b^+ + \cosh\theta} U_{11} + \frac{(-i)^{s+1}}{g_a^+ + \cosh\theta} U_{22}\right) .
\end{equation}
Inserting \cref{eqn:fs4saddlepoint} into the exponential function and expanding in the anticommuting variables and rewriting terms containing $\kappa_0$ as derivatives of $\exp(-ik\kappa_0/16)$ with respect to $k$ yields \cref{eqn:fs4anticommintegration}.

\section{Efetov-Wegner Terms}
\label{app:efetovwegner}
Efetov-Wegner terms occur in superspace if variables are transformed in a way that involves mixing of commuting and anticommuting degrees of freedom. If the integration is not compactly supported, \textit{i.e.} the function does not vanish at the boundaries of the domain, additional surface contributions arise together with the Berzinian of the transformation as first shown in \cite{rothstein1987}. Alternatively, we understand these contributions as a consequence of non-vanishing nilpotent parts that is if we apply a transformation that mixes commuting and anticommuting variables, $x\myprime=f(x,\mu,\mu^\star)$, it is clear that any function of $x$ must be expanded in $\mu$ and $\mu^\star$ and we can not simply replace $x$ by $x\myprime$. This approach also gives the correct surface contributions which is discussed in great detail in Ref. \cite{wegner2016supermathematics}. In general, such calculations are very lengthy and cumbersome, but in this situation, the Efetov-Wegner term reduces to a rather simple form. 

We start at the non-linear sigma model found in \cref{eqn:nonlinearsigma} and observe that the integral for $k=0$ is
\begin{align}\label{eqn:charfctatzero}
	R(0) =& \frac{\mathcal{N}}{2} \int_0^\infty\limits\dd\theta \int_{0}^{\pi}\limits\dd\theta_1 \int_{0}^{\pi}\limits\dd\theta_2 \int_0^{1}\limits\dd u \int_0^{2\pi}\limits\dd\varphi_1 \int_0^{2\pi}\limits\dd\varphi_2 \int_0^{2\pi}\limits\dd\phi_1 \int_0^{2\pi}\limits\dd\phi_2  \notag\\
	&\times \int\dd[\Upsilon]\frac{2u\sin\theta_1 \sin\theta_2 \sinh^3\theta}{\left(\cos\left(\theta_1-\theta_2\right)-\cosh\theta\right)^2 \left(\cos\left(\theta_1+\theta_2\right)-\cosh\theta\right)^2} \notag\\
	&\times \prod_{c=1}^{M} \frac{\left(g_c +\cos\left(\theta_1-\theta_2\right)\right)\left(g_c +\cos\left(\theta_1+\theta_2\right)\right)}{\left(g_c +\cosh\theta\right)^2} ,
\end{align}
where $\dd[\Upsilon]$ is the volume element of all anticommuting variables $\mu_j, \nu_j$ and their complex conjugates. An additional factor $1/2$ occurs, as we have doubled the size of the $\theta_2$ interval, \textit{cf.} \ref{app:parametrisation}. Only considering the ordinary part of the commuting variables and neglecting even orders in the anticommuting variables, we find that the integral must vanish, as no anticommuting variables appear. At the same time, we also see that the Berezinian has poles at $\theta=0$ and $\theta_1-\theta_2=\{0, \pi\}$ and/or $\theta_1+\theta_2=\{0, \pi\}$, of which only the pole $\theta=0$, $\theta_1-\theta_2=\theta_1+\theta_2=\{0, \pi\}$ is an actual pole. The other two cases, where $\theta_1-\theta_2=\{0, \pi\}$ or $\theta_1+\theta_2=\{0, \pi\}$, constitute removable singularities. Let us quickly show that this is the case. We restrict the integral to a sufficiently small region of the domain in the vicinity of the pole and set $\sinh\theta = t \cos\omega, \sin\left(\theta_1+\theta_2\right)= t \sin\omega$ or $\sin\left(\theta_1-\theta_2\right)= t \sin\omega$, respectively. Thus, the full integral behaves as $t^0=1$ for $t$ close to zero and the singularity is removed. Only if both factors in the numerator vanish does a $1/t$-type singularity exist. Furthermore, following the steps of \ref{app:parametrisation}, it is clear that the anticommuting variables are not defined at the origin of the coset space, \textit{i.e.} $Q=-i L$, which corresponds to the point at which the singularities occur, \textit{cf.} the discussion in Ref. \cite{Zirnbauer1986} for a different coset space. Evidently, disregarding even orders in the anticommuting variables is not permissible, and we need to account for the lack of compact support in the fashion illustrated above.

It turns out that the correct treatment proves rather difficult, using the variables $\theta, \theta_1, \theta_2$ and instead $\mu, \mu_1, \mu_2$, \textit{cf.} \ref{app:parametrisation}, are much more suitable. Hence, we transform the integrand and obtain 
\begin{align}
	R(0) =& 2\frac{\mathcal{N}}{2} \int_0^\infty\limits\dd\mu \int_{0}^{1}\limits\dd\mu_1 \int_{\mathclap{-\mu_1}}^{\mathclap{\mu_1}}\limits\dd\mu_2 \int_0^{1}\limits\dd u \int_0^{2\pi}\limits\dd\varphi_1 \int_0^{2\pi}\limits\dd\varphi_2 \int_0^{2\pi}\limits\dd\phi_1 \int_0^{2\pi}\limits\dd\phi_2  \notag\\
	&\quad\times \int\dd[\Upsilon]\frac{4 u}{\sqrt{1-\mu_1^2}\sqrt{1-\mu_2^2}} \frac{\mu^3 \left(1+\mu^2\right) \left(\mu_1^2 - \mu_2^2\right)}{\left(\mu_2^2 + \mu_1^2\right)^2 \left(\mu_2^2 + \mu_2^2\right)^2} \notag\\
	&\quad\times \prod_{c=1}^{M} \frac{\left(g_c + 1 - 2 \mu_2^2\right)\left(g_c + 1 - 2 \mu_1^2\right)}{\left(g_c +1 + 2 \mu^2\right)^2} .
\end{align}
It is crucial to note that the functions $\sin\left(\theta_1 + \theta_2\right)$ and $\sin\left(\theta_1 - \theta_2\right)$ are not independent of each other. Thus, the domain of $\mu_2$ depends on $\mu_1$. The domain of $\mu_1$ splits into the two sets $[0,1]$ and $[1,0]$, as the integrals coincide, we simply restrict ourselves to the former and account for the second domain by a factor $2$. We restrict $\mu_2$ to positive values, as it only appears in the form of $\mu_2^2$ in the integral. We introduce another factor of $2$ to account for this restriction
\begin{align}\label{eqn:CFEfetovWegner}
	R(0) =& 2\mathcal{N} \int_0^\infty\limits\dd\mu \int_{0}^{1}\limits\dd\mu_1 \int_{0}^{\mu_1}\limits\dd\mu_2 \int_0^{1}\limits\dd u \int_0^{2\pi}\limits\dd\varphi_1 \int_0^{2\pi}\limits\dd\varphi_2 \int_0^{2\pi}\limits\dd\phi_1 \int_0^{2\pi}\limits\dd\phi_2  \notag\\
	&\quad\times \int\dd[\Upsilon]\frac{4 u}{\sqrt{1-\mu_1^2}\sqrt{1-\mu_2^2}} \frac{\mu^3 \left(1+\mu^2\right) \left(\mu_1^2 - \mu_2^2\right)}{\left(\mu_2^2 + \mu_1^2\right)^2 \left(\mu_2^2 + \mu_2^2\right)^2} \notag\\
	&\quad\times \prod_{c=1}^{M} \frac{\left(g_c + 1 - 2 \mu_2^2\right)\left(g_c + 1 - 2 \mu_1^2\right)}{\left(g_c +1 + 2 \mu^2\right)^2} .
\end{align}
This integral is exactly the same as in Ref. \cite{VWZ1985}, besides the domains of $\mu, \mu_1$ and $\mu_2$, whose roles are swapped, and the channel factor, which contains twice as many channels due to the additional spin degree of freedom. Due to the invariance of the measure and the channel factor under transformations $T\in \text{UOSp}(2,2\vert 4)$, we can apply a theorem proven by Wegner \cite{Wegner2016}. A direct consequence of this theorem is that, for an invariant function $f(Q)$, with some additional properties, which we do not specify and are fulfilled in this case, we have
\begin{equation}\label{eqn:WegnerIntegral}
	\int\dd\mu(Q) f(Q) = f(-i L) ,
\end{equation}
see Ref. \cite{Zirnbauer1986} for details. Explicit calculations of such integrals for the \emph{GUE} are carried out in Ref. \cite{Wegner2016}, and the generalisation to integration involving the non-compact orthosymplectic supergroup $\text{UOSp}(2,2\vert 4)$ is sketched in Ref. \cite{VWZ1985}. As noted by Ref. \cite{Zirnbauer1986}, the integral in \cref{eqn:WegnerIntegral}, in general, contains additional contributions from the boundaries of integration. However, due to the specific form of $\dd\mu(Q)$ this is not the case here. It is clear that for $\mu\to\infty$ the integrand drops off sufficiently fast, such that all contributions vanish. Similarly, inspecting the upper boundary of $\mu_2$, we see that if we set $\mu_2=\mu_1$ the integral always vanishes as well. Hence, no other contributions besides $\mu=0=\mu_1=\mu_2$ exist, which is exactly the case $Q(\mu=0,\mu_1=0,\mu_2=0)=-i L$. Applying \cref{eqn:WegnerIntegral} to \cref{eqn:CFEfetovWegner}, we find 
\begin{equation}
	R(0) = 2 \mathcal{N} .
\end{equation}
The characteristic function is \textit{per definitionem} normalised to unity at the origin, such that 
\begin{equation}
	R(0) = 2 \mathcal{N} = 1 \Leftrightarrow \mathcal{N} = \frac{1}{2}.
\end{equation}
Thus, we determine the normalisation $\alpha$, and obtain the same Efetov-Wegner term as for the \emph{GUE} and \emph{GOE}, see Ref. \cite{NKSG2014}. 

Finally, we are in the position to properly explain, why it is permissible to only consider the ordinary parts of $\dd\widetilde{M}$ and $\dd\widetilde{\phi}_j$ in \cref{app:volumelemegoldstone}. Applying the same reasoning as above, we find that no Efetov-Wegner terms emerge, as the integral either depends on the angles in a $2\pi$-periodic fashion or does not depend on them at all. Hence, the surface contributions always cancel. 

Besides the Efetov-Wegner term, the only relevant parts for the full characteristic function are those depending only on the ordinary parts of the commuting variables, \textit{cf.} Ref. \cite{Zirnbauer1986}. This leads us to our final expression in \cref{eqn:finalresultmatelem}.

\section{Representations of the Supergroup $\text{OSp}(2n\vert 2n)$ and its Fixed Points}
\label{app:grouptheorygoegse}
The superalgebra $\mathfrak{osp}(2n\vert 2n)$ generating the supergroup $\text{OSp}(2n\vert 2n)$ is defined as the space of supermatrices fulfilling
\begin{equation}\label{eqn:gmet}
	X^\trans g + g X = 0 
\end{equation}
with a metric $g$ whose restriction to the bosonic, fermionic subspace is either the orthogonal or symplectic metric. There exists two different representations
\begin{equation}
	g_1 = \begin{bNiceMatrix}
		X \otimes \mathds{1}_n & 0 \\
		0 & Y \otimes \mathds{1}_n
	\end{bNiceMatrix} \quad \text{and} \quad g_4 = \begin{bNiceMatrix}
	Y \otimes \mathds{1}_n & 0 \\
	0 & X \otimes \mathds{1}_n
	\end{bNiceMatrix}
\end{equation}
which following from \cref{sec:connectiongoegse} correspond to a system without or with integer spin and a system with half-integer spin, respectively. Hence, we suggestively named the two different representations according to the corresponding Dyson index. Clearly, there exists an isomorphic map between the two representations. To derive the representations of the non-compact $\text{UOSp}(2,2\vert 4)$ we need to restrict the algebra to non-compact transformations. This is the subalgebra defined by the fixed-points of the automorphism
\begin{equation}\label{eqn:automorph}
	\chi_\beta\left(x\right) = g_\beta \widetilde{L} x^\star \left(g_\beta \widetilde{L}\right)^{-1} .
\end{equation}
Indeed, it is easy to check that the conditions for the algebra elements in \cref{eqn:gmet,eqn:automorph} are equivalent to the conditions for the transformations $T$ given in \cite{VWZ1985} for $\beta=1$ and in \cref{eqn:tcondition} for $\beta=4$. In summary, the transformations $T$ encoding the symmetries of $B$ for $\beta=1,4$ arise from the fixed-points of two different representations of the orthosymplectic supergroup defined by their respective behaviour under time reversal.

\section*{References}

\bibliography{bibliography_2.bib,Lett_Bib.bib,streu_refs_combined.bib}
\bibliographystyle{iopart-num}

\end{document}